\documentclass[10pt,aps,pra,showpacs,twocolumn,superscriptaddress,longbibliography]{revtex4-2}
\usepackage{graphicx}
\usepackage{subfigure}
\usepackage{subfigure}
\usepackage{amsfonts}
\usepackage{amsmath}
\usepackage{bm}
\usepackage{url}
\usepackage{amssymb}
\usepackage{xcolor}
\usepackage[percent]{overpic}
\usepackage[all,pdf]{xy}
\usepackage[colorlinks,linkcolor=blue]{hyperref}
\makeatletter
\begin{document}
\title{A reinterpretation of classical magnetism via the regular representation of displacement current}
\author{Jin Jer Huang}
\email{huangjinzhe@tiangong.edu.cn}
\affiliation{Department of applied Physics,~Tiangong University, Tianjin 300387, China}

\date{\today}

\begin{abstract}
The displacement current, introduced by Maxwell, has long caused confusion regarding its relationship to the magnetic field. To develop a new understanding of classical magnetism, this work first decomposes the displacement current into a localized internal component and an external field distribution by employing a discal regularization of the dipole distribution. Due to a surprising cancellation of the electric current by this internal displacement current, an integral expression for the magnetic field in terms of the external displacement current (or total current) enables a reinterpretation of classical magnetism. Consequently, the current equivalence is resolved and refined through this integral representation in both quasi-steady and quasi-static states. In the latter case, the nonzero external displacement current is exactly equal to---and can easily be mistaken for---the current in the circuit system. This paper proposes a potential solution to several longstanding historical disputes concerning the displacement current.
\end{abstract}
\maketitle

\section{Introduction}
Since the publication of Maxwell's masterpiece, ``\textit{A Treatise on Electricity and Magnetism}," in 1873 \cite{Maxwell}, which announced the birth of unified electromagnetic theory, one and a half centuries have passed. Remarkably, Maxwell's equations remain the core of classical field theory despite the boom in new physics in the 20th century, except that the microscopic conceptualisation pertinent to the structure of matter has to be modernized. The Maxwell's displacement current is undoubtedly a brilliant innovation that prompted the prediction and discovery of the electromagnetic nature of light, and how this conception emerged from theoretical completeness can be found in most present textbooks \cite{Jackson1,Griffiths,Purcell,Greiner}. However, the old-established definition of the displacement current based on the ether has been mired in controversy for a long time, and thus has become a commemorative phrase in the post-ether era. Nevertheless, the controversy was not about the anachronistic nomenclature of ``displacement current," but rather about its involvement in magnetic field induction. Even now, the theory has led to varying standpoints in education due to disagreements over the understanding of the displacement current. Some valuable reviews can be found in Refs. \cite{Roche,Selvan,Arthur}.

The root of most disputes is from the Amp\`{e}re-Maxwell law, in which the electric current and the displacement current are added together. It means that they share the same physical features to some extent. Not surprisingly, Maxwell proposed a simple equivalent relationship, asserting that ``the displacement current is electromagnetically equivalent to an electric current," thus assuring that every total electric current must have a closed form \cite{Maxwell,Roche}. Maxwell's current equivalence conveys two meanings: 1. The displacement current or total current represents a real electric current, indicating the movement of electricity; 2. The displacement current serves the same role in induction as the conduction current. Point 1 reflects Maxwell's physical reality of the time variation of an electric field as a ``current," but according to modern knowledge of electricity, this statement is not applicable in vacuum \cite{Jackson2}. As for Point 2, it lies at the heart of the controversy and is closely related to the Biot-Savart law. In particular, this equivalence is prone to divergence when each current is utilized as an independent source to induce a magnetic field, partly due to the semantic uncertainty of the term ``source." The striking difference between the two currents prompts us to ask: what exactly is a source? A narrow definition of a source focuses on charge and current, it can be extended to include polarization and magnetization sources, as well as the entities that provide these sources, such as batteries and the ground \cite{Jackson3}. This definition is actually applicable to the concept of an ``ordinary source," which Heras defined as a local, physically realizable, and non-self-referential entity, such as the real current \cite{Heras1}. Jefimenko regarded this kind of source as a primary cause, or origin \cite{Jefimenko1}. Sometimes, it is referred to as an independent, external, or ultimate source. From this perspective, the displacement current cannot be considered as any source of magnetic fields. However, a relaxed definition that does not limit itself to matter is necessary to include field sources. For example, the transverse current is a form of field that acts as a source of the Coulomb vector potential \cite{Jackson1} (Sec. 6.3). In addition, as pointed out by Hill \cite{Hill1,Hill2}, it is conventional in physics education to phrase Faraday's and Amp\`{e}re-Maxwell's laws using ``induction" or ``production", which inherently involves in the concept of source. Indeed, the vortex source has already been established as a fundamental concept \cite{Cheng}. In the electromagnetic field theory, a generalized definition of source, should include the vorticity of an electromagnetic field, setting aside historical debates on causality \cite{Hill1,Hill2,Kinsler1,Savage,Kinsler2}. Moreover, we can classify a cause-and-effect relationship as either direct (local) or indirect. A direct relationship is confined to an immediate and localized causal agent, whereas an indirect relationship pertains to the correlation between the initial and final phenomena in a sequence of events. Direct or local sources arise from the continuous segmentation of an ongoing event. The source can be causal, either direct or indirect, or it may not be a cause at all. In this way, the electric current is undoubtedly an indirect source of a distant electromagnetic wave; however, whether it can serve as a direct source remains a question. Following the convention, some words such as generation, production, and induction are still used here, they, nevertheless, may not indicate an unambiguous cause or effect unless specifically pointed out in this paper. A mathematical foundation of the vortex-field source lies in the Helmholtz decomposition theorem \cite{Jackson2,Miller}, which guarantees that a divergence-free field can be expressed as the curl of the integral of its vortex source. Naturally, this theorem also serves as the basis for the Biot-Savart law and its generalized form. Unfortunately, verifying local causality still lacks robust theoretical support. The retarded effect remains the only available criterion for causality, applicable only to events occurring over a measurable time interval.

Although both the electric current and displacement current are widely recognized as causal sources of magnetic field induction in most textbooks, doubts about the role of the displacement current have persisted for a long time. The skepticism regarding the role of displacement current initially stemmed from the observation that it does not contribute to the magnetic field according to the Biot-Savart law during the steady charging or discharging of a plate capacitor \cite{Roche,Selvan,Arthur,French1,Sciamanda,Heald,Milsom,Hyodo,McDonald2}. This phenomenon has been extensively investigated both experimentally and theoretically \cite{Purcell,Milsom,Hyodo,McDonald2,French2,Zhu,Dahm,Bartlett1,Bartlett2}, which can be attributed to the vanishing curl of the displacement current or the electric field inside, as applicable to any steady circuit system. Readers can find more details in the documents referenced in Ref. \cite{McDonald2}. As for the propagation of electromagnetic waves, some physics educators, such as Rosser \cite{Rosser1,Rosser2} and Arthur \cite{Arthur}, have argued that field sources do not produce electromagnetic fields; instead, they are ultimately caused by external sources. This perspective is based on the retardation mechanism derived from the Lorenz gauge and the wave equation of the Lorenz vector potential. Clearly, they did not believe that the displacement current could exhibit the same mechanism in magnetic field induction as the electric current does, despite their mathematical equivalence. Zapolsky defended Maxwell's displacement current by arguing that Rosser's critique associated with the vector potential is largely semantic, as the displacement current is determined by the divergence-free component of the time derivative of a vector potential and is indirectly incorporated into the potential equation \cite{Zapolsky}. In fact, the relationship between the two types of current is more subtle than commonly understood.

The review above concerning the conundrum of displacement current reveals a logical dilemma rooted in the equivalence in conventional electromagnetism. We are uncertain whether both electric current and displacement current generate a magnetic effect, a question that has recurred throughout history \cite{Roche,Selvan,Arthur}. Affirming this notion is somewhat contentious. The statement that both sources generate the magnetic field independently was criticized by Landini \cite{Landini}, who referred to the magnetic field as ``a son of two mothers." If this were the case, it would be possible to distinguish the contributions to the magnetic field from the two types of current; nevertheless, such a distinction is, in fact, impossible. For a charging capacitor, you can freely choose an Amperian loop and associate it with a surface that may contain either the conduction current or the displacement current (or both), without affecting the magnetic field's circulation. \cite{Griffiths,Jackson3,Dahm,Bartlett1}. Griffiths took it as a quasi-philosophical question \cite{Griffiths}: ``If you measure $\mathbf{B}$ in laboratory, have you detected the effects of displacement current, or merely confirmed the effects of ordinary currents?" This contradiction can be illustrated by the case of a current element, known as Cullwick's paradox \cite{Roche,Cullwick}, in which Cullwick reminded that the two contributions from the current and the displacement current to the induced magnetic field are mutually exclusive and must not be combined. The interpretative challenge lies not only in explaining the cause-and-effect relationship behind the origin of the magnetic field but also in understanding the physical correlation between the two types of current. Whether the displacement current provides an intrinsic physical explanation that goes beyond the mathematical equivalence \cite{Roche}, or serves merely as a conceptual tool \cite{Arthur}, remains an open question.

The direct origin of the magnetic field necessitates a monistic hypothesis from both methodological and philosophical perspectives. The current-source-only standpoint alleviates the dilemma regarding the causality of classical magnetism, albeit only superficially. In principle, the transmission of electromagnetic waves and the movement of a point charge are more fundamental and clearer than those in macroscopic circuit systems, which, after all, are composed of the former. Acting as an external, indirect source, the current is unlikely to be considered an effective candidate for the direct cause of field conversion in an electromagnetic wave. In the case of a uniformly moving point charge, it was found that the magnetic field induced by a current element is canceled out by the singular component of the displacement current flux when it is located on the surface of a specified Amperian loop \cite{Terry}. This suggests that the magnetic field in a capacitor is possibly determined more by the displacement current, according to the principle of superposition. Thus, in both scenarios, the current operates behind the scenes rather than as a direct participant, while the question of whether the displacement current is the direct source of the magnetic field generation demands an in-depth exploration. Moreover, the conventional interpretation of the displacement current is incomplete, and there is also a lack of reliable experimental evidence on the subject, as summarized in Ref. \cite{Hyodo}. In this paper, the field structure of the displacement current is examined to shed new light on related problems within the controversy.  

In Sec. II, the displacement current of a moving point charge is decomposed into an external component and an internal distribution that opposes the real current, resulting in the total current density forming a field distribution. In the continuum approximation, the total current density is connected with a regularized dipole distribution, which is employed to derive an instantaneous Biot-Savart integral and related results in the Coulomb gauge, as detailed in Sec. III. In Sec. IV, an equivalence between the real current and the static displacement current is validated from the instantaneous Biot-Savart integral. Appendices A, B, C, and D present detailed calculations of integrals. An extensive discussion on the causality of field sources follows in Appendix E.

This paper employs the vector analysis to represent electromagnetic fields and adopts a basic sequential approach to handle generalized functions as specific conditional limits of regular functions (see, for example, Ref. \cite{Jones}). A simple application of the distribution concept is adopted, omitting a modern treatment based on functional or measure-theoretic methods, making it more accessible to common readers. Consequently, the expressions of singular displacement current densities will be denoted with tildes to make a difference from ordinary fields or functions, where they should be used in an integral representation. Besides, all equations in this paper are presented in SI units.

\section{Displacement current of a moving charged particle}
It is assumed that all external sources, such as electric charges and currents, are confined to a compact region in free space. All sources and fields are associated with a space-time point $(\mathbf{x}, t)$, which will be dropped in the following equations unless necessary. A general real electric current, including free conduction, convection currents, and other types of bound currents, will be referred to simply as current below for short. Furthermore, the word ``density" in certain phrases, such as ``current density," will be omitted after its initial appearance if it does not cause ambiguity. Then, the well-known Amp\`{e}re-Maxwell law has the form
\begin{align}
\nabla\times\mathbf{H}=\mathbf{J}+\frac{\partial\mathbf{D}}{\partial t}\,,
\label{A-M-Law}
\end{align}
where $\mathbf{J}$ is the current density, $\nabla$ is the nabla (gradient) operator with respect to the position vector $\mathbf{x}$, $\mathbf{H}$ is the magnetic field related to the magnetic induction $\mathbf{B}$ by $\mathbf{H}$=$\mu_0^{-1}\mathbf{B}$, and $\mathbf{D}$ represents the electric displacement vector (or electric flux density), related to the electric field $\mathbf{E}$ by $\mathbf{D}$=$\epsilon_0\mathbf{E}$. In the above relations, $\epsilon_0$ and $\mu_0$ are the permittivity and the permeability in vacuum, respectively. The last term in Eq.~(\ref{A-M-Law}) is just the displacement current density, whose divergence is
\begin{align}
\nabla\cdot\left(\frac{\partial\mathbf{D}}{\partial t}\right)=-\nabla\cdot\mathbf{J}=\frac{\partial \rho}{\partial t}\,.
\label{A-M-Law-Div}
\end{align}
This is the continuity equation with the charge density $\rho$. 

At the point of a charged particle, the associated charge and current densities carry a singular spatial distribution. In this case, Eq.~(\ref{A-M-Law-Div}) implies that the displacement current also exhibits this singularity because the right-hand side (rhs.) of the equation corresponds to a distribution of charged point sources. Hence, a current composed of moving point charges in a conductor simultaneously generates a displacement current with a highly localized distribution. This singular distribution may be inadvertently overlooked due to the spatial averaging in the continuum approximation used in macroscopic electromagnetism \cite{Jackson1,Russakoff}. It implies that the displacement current inherits the singularity at a macroscopic scale, which will be addressed below. The flux discontinuity of an electric field was first noticed by Rosser and was applied to a moving charged body of an arbitrary shape \cite{Rosser1}. In 1982, Terry found that the electric flux generated by a uniformly moving point charge includes a local distribution that is singular in the direction of motion, resembling a reverse current \cite{Terry}. This local electric flux cancels the current element when calculating an induced magnetic field. 

\begin{figure}[ht]
\centering
\begin{overpic}
[scale=0.65]{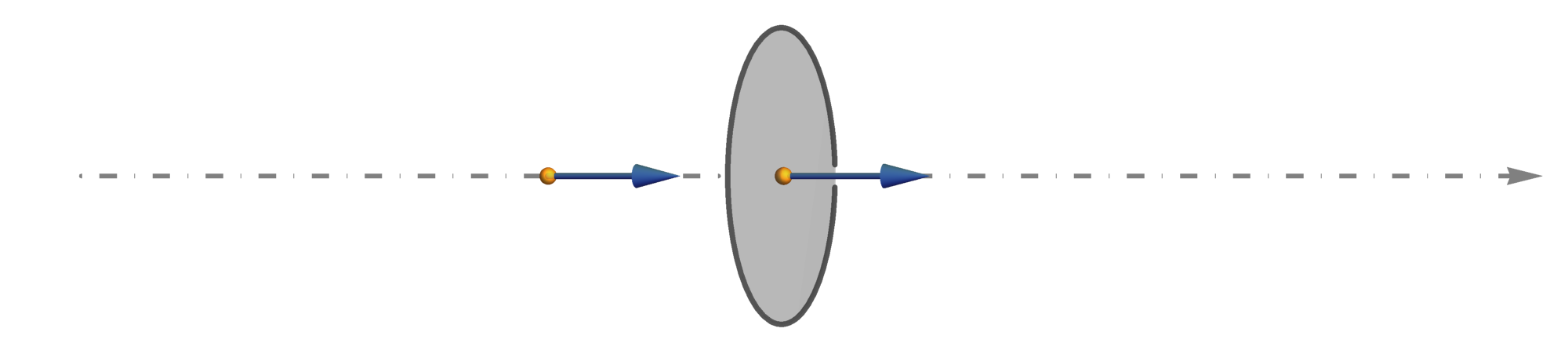}
\put(56,7){$\bm{\upsilon}$}
\put(40,7){$\bm{\upsilon}$}
\put(49,7){0}
\put(49,14){$q$}
\put(54,15){$S$}
\put(98,10.5){$z$}
\put(5,18){(a)}
\put(35,10){\color{gray}\line(0,-1){60.5}}
\end{overpic}
\begin{overpic}
[scale=0.65]{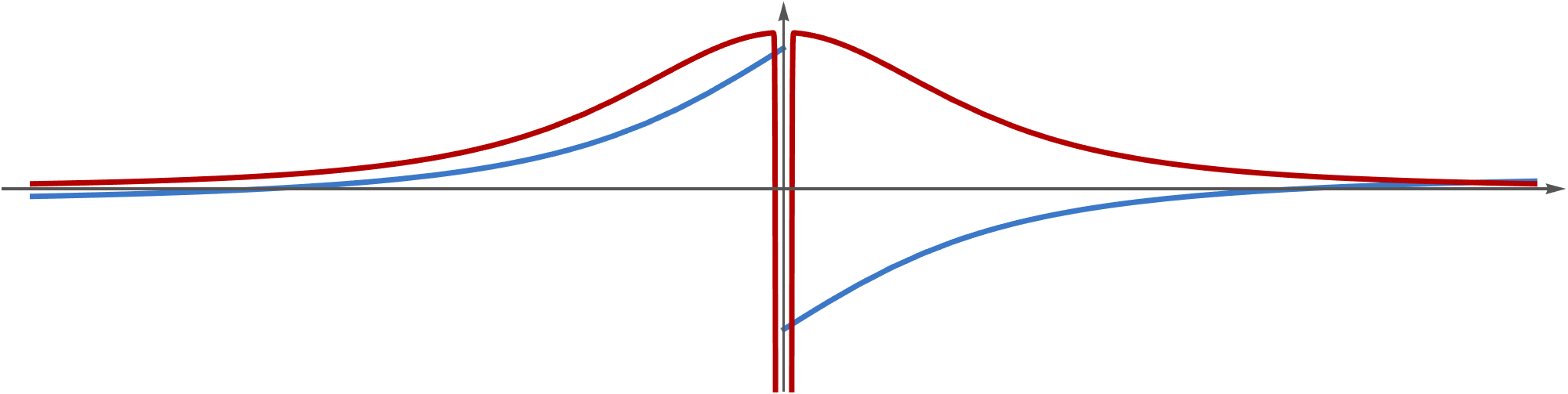}
\put(63,7){$-q\Theta(z)+\Psi$}
\put(60,19){$-q\upsilon\delta(z)+\mathrm{d}\Psi/\mathrm{d}t$}
\put(47,10){0}
\put(98,15){$z$}
\put(5,18){(b)}
\end{overpic}
\begin{overpic}
[scale=0.65]{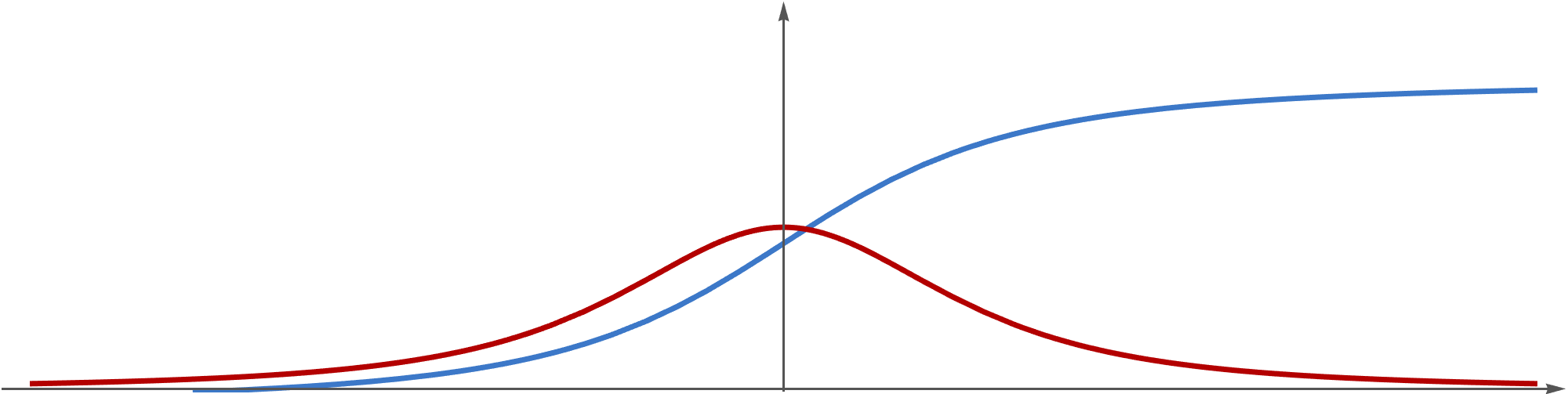}
\put(66.5,13){$\Psi$}
\put(64.5,5){$\mathrm{d}\Psi/\mathrm{d}t$}
\put(47,1){0}
\put(98,2){$z$}
\put(5,18){(c)}
\end{overpic}
\caption{\label{fig-1} (a) Schematic of a particle with charge $q$ moving through the center of a circular surface $S$; (b) plots of corresponding electric flux (blue), displacement current (red); and (c) plots of total current (red) and the nonsingular electric flux (blue) along the moving path on the $z$ axis.}
\end{figure}

Let's revisit this enlightening example of a point charge of $q$ moving at a constant velocity $\bm{\upsilon}$. The symmetric geometry, electric flux, and displacement current of the point charge moving along the $z$ coordinate are shown in Fig. \ref{fig-1}. The electric flux through a fixed circular surface $S$ can be written as $-q\Theta(z)$$+$$\Psi$, where $\Psi$ is a nonsingular part having the form of $\tfrac{1}{2}q(1+z/r)$ with a distance $r$ from the particle to the edge of $S$ \cite{Terry}. The singular components therein, as shown in Fig. \ref{fig-1}(b), appear as the Heaviside function $\Theta(z)$ in the flux and the Dirac function $\delta(z)$ in the displacement current, respectively. However, in the expression for the total current, the singular terms are cancelled out, illustrated in Fig. \ref{fig-1}(c). In the case of a current line model, Weber and Macomb also confirmed a current cancellation in presenting the displacement current \cite{Weber}. A followed work by Liang and Wang verified the current cancellation again \cite{Liang}, and it was explained as the conservation of charge (not for density): ``the conduction current plus the displacement current is zero." The problem of a charged, moving point particle will be further examined in the following subsections.

\subsection{The nonrelativistic case}
In the following analysis, it is assumed that a particle of charge $q$ moves with velocity $\bm{\upsilon}$ satisfying $|\bm{\upsilon}|$$\ll$$c$ in the observer's (laboratory) rest frame, where $c$ is the speed of light. Therefore, relativistic effects will not be considered here. Let $r$ be the distance from the observer's point $\mathbf{x}$ to the location $\mathbf{x}^\prime$ of the particle at time $t$. Then, the potential and displacement vector can be written as $\varphi_q$=$q/(4\pi\epsilon_0 r)$ and $\mathbf{D}_q$=$-\epsilon_0\nabla\varphi_q$, respectively. Thus, the related displacement current can be generally written as
\begin{align}
&\left(\frac{\partial\mathbf{D}_q}{\partial t}\right)_{\!\!r>0}=\left.\bm{\upsilon}\cdot(\nabla^{\prime}\mathbf{D}_q)\right|_{r>0}=\left.\bm{\upsilon}\cdot(\nabla\nabla\epsilon_0\varphi_q)\right|_{r>0}\nonumber\\
&=\left.\frac{q}{4\pi}\bm{\upsilon}\cdot\nabla\left(\frac{-\mathbf{r}}{r^3}\right)\right|_{r>0}
=\frac{q}{4\pi r^3}(3\bm{\upsilon}\cdot\mathbf{\hat{r}}\mathbf{\hat{r}}-\bm{\upsilon})\,,
\label{Nonlocal-DC}
\end{align}
where the prime symbol on the nabla operator denotes the derivative with respect to the source point $\mathbf{x}^{\prime}$, and $\mathbf{r}$=$\mathbf{x}$$-$$\mathbf{x}^{\prime}$ is a spatial displacement. The hatted vector $\mathbf{\hat{r}}$ in Eq.~(\ref{Nonlocal-DC}) denotes the unit vector of $\mathbf{r}$, a convention that is used throughout this paper.
\begin{figure}[ht]
\centering
\begin{overpic}
[scale=0.8]{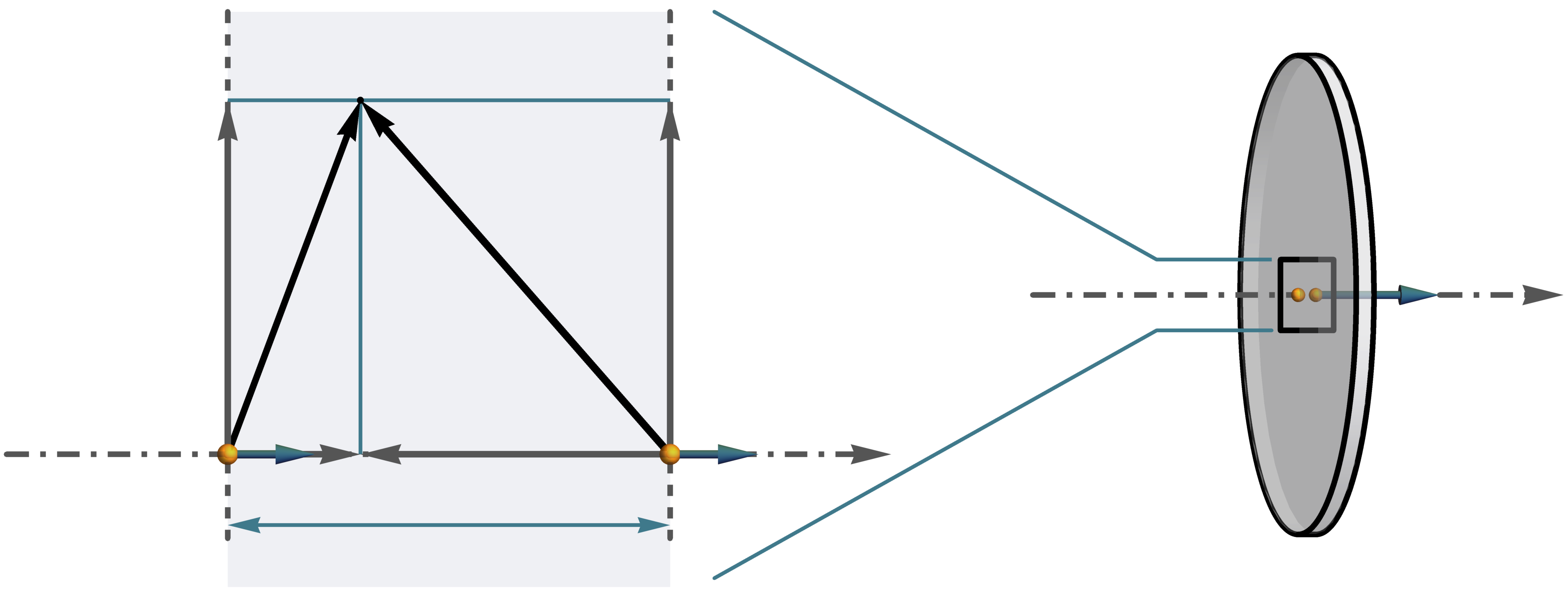}
\put(11,11){$\mathbf{x}^{\prime}$}
\put(12,6.5){$q$}
\put(10,28){$\mathbf{r}_{\perp}$}
\put(44,28){$\mathbf{R}_{\perp}$}
\put(44,11){$\mathbf{x}^{\prime\prime}$}
\put(40,6.5){$q$}
\put(46,6){$\bm{\upsilon}$}
\put(17,6){$\bm{\upsilon}$}
\put(22,33){$\mathbf{x}$}
\put(18,25){$\mathbf{r}$}
\put(30,25){$\mathbf{R}$}
\put(30,11){$\mathbf{R}_{\parallel}$}
\put(19,11){$\mathbf{r}_{\parallel}$}
\put(13,1){$t_1$}
\put(41,1){$t_2$}
\put(22,1){$\Delta t, r^{\prime}$}
\put(72,36){disk: $V^{\prime}_\mathrm{d}$}
\put(92,19.5){\color{gray}\vector(0,-1){15.2}}
\put(92,4.3){\color{gray}\vector(0,1){15.2}}
\put(86,4.3){\color{gray}\line(1,0){7}}
\put(88.5,11){$r_{\mathrm{d}}$}
\end{overpic}
\caption{\label{fig-2} Schematic of a disk-like cylinder along the moving direction of a charged particle, formed by two circular cross-sectional surfaces at two moments of motion.}
\end{figure}
Equation (\ref{Nonlocal-DC}) represents a $\bm{\upsilon}$-axis symmetrical vector field that is singular at $\mathbf{x}^{\prime}$, analogous to the field of a dipole. According to the aforementioned discussion, the displacement current is meaningful at $\mathbf{x}^{\prime}$ as a distribution but is not well-defined and therefore requires a regularization. However, the regularized result is not unique \cite{Hnizdo}. In a simple way, we can start from a most intuitive approach, by simply returning to the definition of the derivative as the limit of a difference quotient. During the movement of the point charge within a quite small time interval $\Delta t$, its location will shift from $\mathbf{x}^{\prime}$ to $\mathbf{x}^{\prime\prime}$ with a displacement $\mathbf{r}^{\prime}$=$\mathbf{x}^{\prime}$$-$$\mathbf{x}^{\prime\prime}$, which moves $\mathbf{r}$ to $\mathbf{R}$=$\mathbf{x}$$-$$\mathbf{x}^{\prime\prime}$, as shown in Fig. \ref{fig-2}. The geometrical variables therein are set as $r^{\prime}$=$|\mathbf{r}^{\prime}|$, $r_{\parallel}$=$\mathbf{r}\cdot\bm{\hat{\upsilon}}$, $\mathbf{r}_{\parallel}$=$r_{\parallel}\bm{\hat{\upsilon}}$, $\mathbf{R}_{\parallel}$=$-R_{\parallel}\bm{\hat{\upsilon}}$, and $\mathbf{r}_{\perp}$=$\mathbf{r}$$-$$\mathbf{r}_{\parallel}$=$\mathbf{R}_{\perp}$. We can delineate an internal region $V^{\prime}_\mathrm{d}$, a small flat cylinder situated between two circular surfaces with a radius $r_{\mathrm{d}}$, normal to the direction of motion, i.e.,
\begin{align}
V^{\prime}_\mathrm{d}=\{\mathbf{x}| r_{\perp}<r_{\mathrm{d}}, 0\leqslant r_{\parallel}\leqslant r^{\prime}>0\}
\label{Disc-region}
\end{align}
with $r^{\prime}$$\ll$$r_{\mathrm{d}}$. The prime symbol in $V^{\prime}_\mathrm{d}$ indicates the point $\mathbf{x}^{\prime}$ at which the disk is situated.  
For convenience, $r_{\mathrm{d}}$ can be arbitrarily small, which compresses $V^{\prime}_\mathrm{d}$ into an infinitesimal disk. In this way, the singular internal component (denoted by the superscript ``int") of the displacement current can be taken as the limit of a difference quotient within $V^{\prime}_\mathrm{d}$
\begin{align}
\left(\frac{\partial\mathbf{D}_q}{\partial t}\right)^{\!\!\mathrm{int}}=&\lim_{\Delta t\rightarrow 0}\mathrm{rect}\left(\frac{r_{\parallel}}{r^{\prime}}\right)\frac{\mathbf{D}_q(\mathbf{R})-\mathbf{D}_q(\mathbf{r})}{\Delta t}\nonumber\\
=&\upsilon\lim_{r^{\prime}\rightarrow 0}\mathrm{rect}\left(\frac{r_{\parallel}}{r^{\prime}}\right)\frac{\mathbf{D}_q(\mathbf{R})-\mathbf{D}_q(\mathbf{r})}{r^{\prime}}\,,
\label{Local-DC-Limit-1}
\end{align}
where the rectangular function $\mathrm{rect}$ is defined by
\begin{equation}
\mathrm{rect}(\xi)=
\begin{cases}
1, \xi\in [0,1]\\
0, \xi\notin [0,1]
\end{cases}\,.
\label{Rectangular-Function}
\end{equation}
The difference of the two electric displacement vectors in Eq.~(\ref{Local-DC-Limit-1}) can be factored as
\begin{align}
&\mathbf{D}_q(\mathbf{R})-\mathbf{D}_q(\mathbf{r})=\frac{-q}{4\pi}\left(\frac{\mathbf{r}_{\perp}+\mathbf{r}_{\parallel}}{r^3}
-\frac{\mathbf{R}_{\perp}+\mathbf{R}_{\parallel}}{R^3}\right)\nonumber\\
&=-\frac{q\bm{\hat{\upsilon}}}{4\pi}\left(\frac{r_{\parallel}}{r^3}
+\frac{R_{\parallel}}{R^3}\right)-\frac{q\mathbf{r}_{\perp}}{4\pi}\left(\frac{1}{r^3}
-\frac{1}{R^3}\right)\,.
\label{Difference-D}
\end{align}
Due to the axial symmetry, the average of the transverse components proportional to $\mathbf{r}_{\perp}$ in Eq.~(\ref{Difference-D}) vanishes as the disk-like region $V^{\prime}_\mathrm{d}$ shrinks to a point. Hence, the distributional meaning of Eq.~(\ref{Difference-D}) pertains only to the longitudinal electric flux through a cross-section. Therefore, Eq.~(\ref{Local-DC-Limit-1}) results as
\begin{align}
\left(\frac{\partial\mathbf{D}_q}{\partial t}\right)^{\!\!\mathrm{int}}=&-\lim_{r^{\prime}\rightarrow 0}\mathrm{rect}\left(\frac{r_{\parallel}}{r^{\prime}}\right)\frac{q\bm{\upsilon}}{4\pi r^{\prime}}\left(\frac{r_{\parallel}}{r^3}
+\frac{R_{\parallel}}{R^3}\right)\nonumber\\
=&-\delta(\mathbf{r})q\bm{\upsilon}\,,
\label{Local-DC-Limit-2}
\end{align}
where, the calculation of the limit is shown in Appendix A. 

Interestingly, Eq.~(\ref{Local-DC-Limit-2}) has the same form as the current density but is associated with the moving point charge $-q$. Hence, the displacement current can reasonably be expressed as the sum of its internal and external components, with the latter denoted by the superscript ``ext," i.e.,
\begin{align}
\mathbf{J}_{\mathrm{D}q}=\epsilon_0\bm{\upsilon}\cdot\nabla\nabla\varphi_q=\mathbf{\tilde{J}}^{\mathrm{int}}_{\mathrm{D}q}+\mathbf{J}^{\mathrm{ext}}_{\mathrm{D}q}\,,
\label{DC-q-Case}
\end{align}
where the two components are
\begin{subequations}\label{DC-q}
\begin{align}\label{DC-q-1}
\mathbf{\tilde{J}}^{\mathrm{int}}_{\mathrm{D}q}(\mathbf{x};\mathbf{x^{\prime}}(t))=&-\delta(\mathbf{r})q\bm{\upsilon}\,,\mathbf{r}\in V^{\prime}_{\mathrm{d}}\,,\\
\label{DC-q-2}
\mathbf{J}^{\mathrm{ext}}_{\mathrm{D}q}(\mathbf{x};\mathbf{x^{\prime}}(t))=&
\begin{cases}
\epsilon_0\bm{\upsilon}\cdot\nabla\nabla\varphi_q, \mathbf{r}\in\bar{V}^{\prime}_{\mathrm{d}}\\
0, \mathbf{r}\in V^{\prime}_{\mathrm{d}}
\end{cases}\nonumber\\
=&\,q\left(\frac{3\bm{\upsilon}\cdot\mathbf{\hat{r}}\mathbf{\hat{r}}-\bm{\upsilon}}{4\pi r^3}\right)_{\!\!\bar{V}^{\prime}_{\mathrm{d}}},
\bar{V}^{\prime}_{\mathrm{d}}=V_{\infty}/V^{\prime}_{\mathrm{d}}\,.
\end{align}
\end{subequations}
In Eq.~(\ref{DC-q-2}), $V_{\infty}$ represents all space, and the subscript $\bar{V}^{\prime}_{\mathrm{d}}$ denotes the region obtained by excluding the infinitesimal disk, a convention that will be used in the following context. Actually, Eqs.~(\ref{DC-q}) result from the discal regularization, a special case of cylindrical regularization, as demonstrated in Appendix B. As seen from the above equations, the internal displacement current is just the origin of the singular electric flux illustrated in Fig.~\ref{fig-1}. As a correspondence, the following equations explain the results in Fig.~\ref{fig-1}, i.e.,
\begin{align}\label{Terry-q}
-q\upsilon\delta(z)=\int_S\mathbf{\tilde{J}}^{\mathrm{int}}_{\mathrm{D}q}\cdot\mathrm{d}\mathbf{s}\,,
\frac{\mathrm{d}\Psi}{\mathrm{d}t}=\lim_{r_{\parallel}\rightarrow z}\int_S\mathbf{J}^{\mathrm{ext}}_{\mathrm{D}q}\cdot\mathrm{d}\mathbf{s}\,.
\end{align}

\subsection{The relativistic case}
Even with relativity effects included, the same internal distribution of the displacement current persists, as will be demonstrated below. Based on the Li\'{e}nard-Wiechert potentials, the electric field of a moving point particle, expressed at the retarded time, can be derived and found in many textbooks \cite{Jackson1,Griffiths}, i.e.,
\begin{align}
\mathbf{D}_{q\mathrm{R}}=\frac{q}{4\pi}\frac{r}{(\mathbf{r}\cdot\mathbf{u})^3}[(c^2-\upsilon^2)\mathbf{u}+\mathbf{r}\times(\mathbf{u}\times\mathbf{a})]
\label{Relativity-Q-D-1}
\end{align}
with $\mathbf{u}$$=$$c\mathbf{\hat{r}}$$-$$\bm{\upsilon}$ and an acceleration $\mathbf{a}$. The second part of Eq.~(\ref{Relativity-Q-D-1}) is the acceleration field that is related to the electromagnetic radiation, a nonlocal term. In comparison, it is one order of magnitude less than the first term, which is the velocity field, as the distance $r$ approaches zero. Hence, only the local velocity field, a generalized Coulomb field, is strong enough to form a singular distribution near the source point. In this case, a uniformly moving point charge without acceleration exhibits the same local singularity as one that has the same velocity but possesses a nonzero acceleration. The electric displacement of the charged particle with a constant velocity can be expressed at an instantaneous time $t$ as
\begin{align}
\mathbf{D}_{q\mathrm{R}}=\frac{\gamma q}{4\pi}\frac{\mathbf{r}}{[r^2+(\gamma\mathbf{r}\cdot\bm{\upsilon}/c)^2]^{3/2}}
\label{Relativity-Q-D-2}
\end{align}
where $\gamma$$=$($1$$-$$\upsilon^2/c^2$)$^{-1/2}$. After two symbolic substitutions of $r_{\mathrm{r}}^2$=$r^2_{\perp}$$+$$\gamma^2r^2_{\parallel}$ and $R_{\mathrm{r}}^2$=$R^2_{\perp}+\gamma^2R^2_{\parallel}$, Eq. (\ref{Difference-D}) becomes
\begin{align}
&\frac{-q\gamma}{4\pi}\left(\frac{\mathbf{r}_{\perp}+\mathbf{r}_{\parallel}}{r_{\mathrm{r}}^3}
-\frac{\mathbf{r}_{\perp}+\mathbf{R}_{\parallel}}{R_{\mathrm{r}}^3}\right)\nonumber\\
=&\frac{-q\gamma\bm{\hat{\upsilon}}}{4\pi}\left(\frac{r_{\parallel}}{r_{\mathrm{r}}^3}
+\frac{R_{\parallel}}{R_{\mathrm{r}}^3}\right)-\frac{q\gamma\mathbf{r}_{\perp}}{4\pi}\left(\frac{1}{r_{\mathrm{r}}^3}
-\frac{1}{R_{\mathrm{r}}^3}\right)\,,
\label{Relativity-Difference-Q-D}
\end{align}
where the two terms in the last bracket can be neglected as before. Equation (\ref{Relativity-Difference-Q-D}) still evolves into the Dirac function, which is shown in Appendix A. In this way, Eq.~(\ref{DC-q-1}) still holds. From Eq. (\ref{Relativity-Q-D-2}), the generalized external displacement current can be derived but is omitted here. According to the above results, the displacement current of a continuous current can also be decomposed into two components:
\begin{align}
\mathbf{J}_{\mathrm{D}}=\mathbf{J}_{\mathrm{D}}^{\mathrm{int}}+\mathbf{J}_{\mathrm{D}}^{\mathrm{ext}}\,.
\label{DC-General-Case}
\end{align}
It will be further discussed in the next Sec..

\subsection{Discal regularization}
The generalization of the distribution $\nabla\nabla (1/r)$ to all space is a typical regularization issue \cite{Estrada}. Owing to specific concerns regarding the internal distribution, only the regularization to include $r$$=$0 \cite{Hnizdo}, rather than the external one to $r$=$\infty$ \cite{Petrascheck}, is considered here. The regularization of a dipole field at the source point, typically implemented by introducing a singular distribution at $r$$=$0, provides a mathematically complete description. In fact, its internal distribution has not yet received broad attention due to its limited applications. A notable practical case arises from the field distribution limit of a polarized spherical material, which is used in a microscopic polarization model to account for the differences between the effective field and the average field \cite{Born}. The polarized sphere model incorporates a Dirac $\delta$ function when the sphere is approximated as a point \cite{Jackson1} (Sec. 4.1). As one of the few applications, this internal singularity can be used to deal with the hyperfine interactions \cite{Skinner}. The regularization has gained attention since Frahm's systematic work \cite{Frahm}. According to the expression from Skinner and Weil \cite{Skinner}, the dipole field can be reformulated as a dyadic Frahm formula
\begin{align}
\nabla\nabla\frac{1}{r}=\left(\frac{3\mathbf{\hat{r}}\mathbf{\hat{r}}-\bm{\mathcal{I}}}{r^3}\right)_{\!\!\bar{V}^{\prime}_{\mathrm{s}}}-\frac{4\pi}{3}\delta(\mathbf{r})\bm{\mathcal{I}}
\label{Ball-Regularization}
\end{align}
with the unit tensor $\bm{\mathcal{I}}$. Equation~(\ref{Ball-Regularization}) is written in accord with the common spherical regularization as a tensor-valued distribution, where $\bar{V}^{\prime}_{\mathrm{s}}$ denotes the region with an infinitesimal sphere at $r$$=$0 excluded. The total dipole distribution can be decomposed into two parts: a dipole distribution and a singular distribution, as shown on the rhs. of Eq.~(\ref{Ball-Regularization}); however, it should be noted that this decomposition is not unique. Farassat demonstrated that the regularization of the dipole field function is contingent upon the shape of a designated internal zone \cite{Hnizdo}. Although different total dipole distributions are equivalent to one another \cite{Farassat}, their components are not. For example, Eqs.~(\ref{DC-q}) result from a normal discal regularization, which yields a dyadic total dipole distribution
\begin{align}
\nabla\nabla\frac{1}{r}=\left(\frac{3\mathbf{\hat{r}}\mathbf{\hat{r}}-\bm{\mathcal{I}}}{r^3}\right)_{\!\!\bar{V}^{\prime}_{\mathrm{d}}(\mathbf{\hat{s}})}-4\pi\delta(\mathbf{r})\mathbf{\hat{s}}\mathbf{\hat{s}}\,,
\label{Disc-Regularization}
\end{align}
according to Eqs.~(\ref{Internal-Regularization}), (\ref{D11-Estimation}) and (\ref{Appb-C-Regularization-D33}) in Appendix B, where $\mathbf{\hat{s}}$ denotes the normal direction of the disk therein.

Generally, there are nuanced dissimilarities between two different regularization methods, even when they are based on the same geometry. A specific regularization technique with its corresponding method is meaningful only when it can establish a physical correspondence with geometric similarity. Thus, only the spherical regularization is suitable for hyperfine interactions due to its spherical symmetry. The intuitive regularization of the displacement current provides a general difference-quotient method, which can be implemented by calculating the electric field of a uniformly charged volume $V^{\prime}_{\varepsilon}$ at a fixed boundary point, as elaborated in Appendix B. This method provides the same spherical and discal regularizations as those shown by Hnizdo \cite{Hnizdo}, but it differs in other geometries.

However, the geometric constraint on regularization may pose a problem for its applications. One regularization may not be suitable for computations in a given coordinate system, which might be greatly simplified in another form of regularization. For example, the normal direction $\mathbf{\hat{s}}$ of a specified disk should coincide with the corresponding current direction $\mathbf{\hat{n}}$=$\bm{\hat{\upsilon}}$, as manifested in Eqs.~(\ref{DC-q}). This means that $\mathbf{\hat{s}}$=$\mathbf{\hat{n}}$ is required, but it may not hold in practice. The $\mathbf{\hat{n}}$=$\mathbf{\hat{n}}(\mathbf{x^{\prime}})$ component of a total dipole distribution according to Eq.~(\ref{Internal-Regularization}) can be rewritten as
\begin{align}
\nabla\nabla\frac{1}{r}\cdot\mathbf{\hat{n}}=&\left(\frac{3\mathbf{\hat{r}}\mathbf{\hat{r}}\cdot\mathbf{\hat{n}}-\mathbf{\hat{n}}}{r^3}\right)_{\!\!\bar{V}^{\prime}_{\varepsilon}}
-4\pi\delta(\mathbf{r})\bm{\mathcal{D}}_{\varepsilon}\cdot\mathbf{\hat{n}}\nonumber\\
=&\left(\frac{3\mathbf{\hat{r}}\mathbf{\hat{r}}\cdot\mathbf{\hat{n}}-\mathbf{\hat{n}}}{r^3}\right)_{\!\!\bar{V}^{\prime}_{\varepsilon}}
+4\pi\delta(\mathbf{r})\bm{\mathcal{E}}\cdot\mathbf{\hat{n}}\nonumber\\
-&4\pi\delta(\mathbf{r})\mathbf{\hat{n}}\,,
\label{General-Dn}
\end{align}
where $\bm{\mathcal{E}}$=$\bm{\mathcal{I}}$$-$$\bm{\mathcal{D}}_{\varepsilon}$ is a tensor-valued compensation factor. In Eq.~(\ref{General-Dn}), the total dipole distribution in an arbitrary $\varepsilon$-regularization is rewritten in terms of a specified discal regularization with its disk normal aligned along $\mathbf{\hat{n}}$. Hence, the dipole distribution for the normal discal regularization with $\mathbf{\hat{s}}$=$\mathbf{\hat{n}}$ has an equivalent form
\begin{align}\label{Effective-Dipole-distribution}
\left(\frac{3\mathbf{\hat{r}}\mathbf{\hat{r}}-\bm{\mathcal{I}}}{r^3}\right)_{\!\!\bar{V}^{\prime}_{\mathrm{d}}(\mathbf{\hat{n}})}
\simeq&\left(\frac{3\mathbf{\hat{r}}\mathbf{\hat{r}}-\bm{\mathcal{I}}}{r^3}\right)_{\!\!\bar{V}^{\prime}_{\varepsilon}}
+4\pi\delta(\mathbf{r})\bm{\mathcal{E}}\,,
\end{align}
where ``$\simeq$" represents an equivalence relationship in the meaning of distribution. A regularization transformation will facilitate the application of the dipole distribution. Specifically, for a tilted-disk regularization indicated by $\bar{V}^{\prime}_{\varepsilon}$=$\bar{V}^{\prime}_{\mathrm{d}}(\mathbf{\hat{s}})$, the equivalence yields $\bm{\mathcal{E}}$=$\bm{\mathcal{I}}$$-$$\mathbf{\hat{s}}\mathbf{\hat{s}}$, which will be applied in the next Sec.. The external displacement current density in Eq.~(\ref{DC-q-2}) can thus be rewritten in an arbitrary $\varepsilon$-regularization as
\begin{align}\label{DC-External-Distribution}
\mathbf{\tilde{J}}^{\mathrm{ext}}_{\mathrm{D}q}(\mathbf{x};\mathbf{x^{\prime}})
=\,q\left(\frac{3\bm{\upsilon}\cdot\mathbf{\hat{r}}\mathbf{\hat{r}}-\bm{\upsilon}}{4\pi r^3}\right)_{\bar{V}^{\prime}_{\varepsilon}}+\delta(\mathbf{r})\bm{\mathcal{E}}\cdot q\bm{\upsilon}\,.
\end{align}
Equation~(\ref{DC-External-Distribution}) has an exchange symmetry between $\mathbf{x^{\prime}}$ and $\mathbf{x}$. 

The discal regularization transforms the projection operators into effective distributions by allocating a disk-shaped internal region. However, the internal field boundaries cannot be physically determined, nor can any alternative definition be excluded. In fact, other regularizations, as shown in Appendix B, can be interpreted as stipulating the shape of the interior of a dipole. Nevertheless, these alternative definitions do not provide a comprehensible form of the total current density $\mathbf{J}_{\mathrm{tot}}$=$\mathbf{J}$$+$$\mathbf{J}_{\mathrm{D}}$. It will be seen later that the normal discal regularization is essential not only for the displacement current but also for a general field decomposition.

\section{Applications in the Coulomb gauge}
The continuum approximation, assuming a continuous distribution of charges and currents, is applied below. In this Sec., the vector potential and the external displacement current in the Coulomb gauge are derived from the wave equations of the potentials to facilitate further development, aided by the dipole distribution function.
\subsection{Vector potential in the Coulomb gauge}
The two components of an instantaneous displacement current can be found by integrating Eqs.~(\ref{DC-q-1}) and (\ref{DC-External-Distribution}), respectively, after substituting $\rho\mathrm{d}^3x^{\prime}$ for $q$ and interchanging $\mathbf{x^{\prime}}$ and $\mathbf{x}$; i.e.,
\begin{subequations}\label{General-Continous-Nonlocal-DC}
\begin{align}\label{General-Continous-Nonlocal-DC-1}
\mathbf{J}^{\mathrm{int}}_{\mathrm{D}}=&-\mathbf{J}\,,\\
\mathbf{J}^{\mathrm{ext}}_{\mathrm{D}}=&\,\bm{\mathcal{E}}\cdot\mathbf{J}+\int_{\bar{V}_{\varepsilon}}
\!\frac{(3\mathbf{\hat{r}}\mathbf{\hat{r}}-\bm{\mathcal{I}})\cdot\mathbf{J}(\mathbf{x}^{\prime})}{4\pi r^3}\mathrm{d}^3x^{\prime}\,.
\label{General-Continous-Nonlocal-DC-2}
\end{align}
\end{subequations}
Equation (\ref{General-Continous-Nonlocal-DC-2}) stands for a static external displacement current without retardation. As $\mathbf{x}^{\prime}$ becomes an integration variable, the integration region $\bar{V}^{\prime}_{\varepsilon}$ turns into $\bar{V}_{\varepsilon}$, which is allowed by the exchange symmetry in Eq.~(\ref{DC-External-Distribution}).

The total dipole distribution can also be decomposed as operator representations by using the basic operator identity \cite{Jackson1,Greiner}
\begin{align}
-\nabla^2=(\nabla\times\nabla\times)-(\nabla\nabla\cdot)\,.
\label{Laplacian-Decomposition}
\end{align}
Two (integral) operators are commonly defined as
\begin{subequations}\label{Operators}
\begin{align}\label{Longitudinal-Operator}
\bm{\mathcal{L}}_{\varepsilon}(\mathbf{x},\mathbf{x}^{\prime})\equiv&-\nabla\nabla\frac{1}{4\pi r}
=\delta(\mathbf{r})\bm{\mathcal{D}}_{\varepsilon}-\bm{\mathcal{O}}_{\varepsilon}(\mathbf{x},\mathbf{x}^{\prime})\,,\\
\label{Transverse-Operator}
\bm{\mathcal{T}}_{\varepsilon}(\mathbf{x},\mathbf{x}^{\prime})\equiv&\,\nabla\times\nabla\frac{1}{4\pi r}(\times)=\left(-\nabla^2+\nabla\nabla\right)\frac{1}{4\pi r}\nonumber\\
=&\,\delta(\mathbf{r})\bm{\mathcal{I}}-\bm{\mathcal{L}}_{\varepsilon}
=\delta(\mathbf{r})\bm{\mathcal{E}}+\bm{\mathcal{O}}_{\varepsilon}(\mathbf{x},\mathbf{x}^{\prime})\,,\\
\bm{\mathcal{O}}_{\varepsilon}(\mathbf{x},\mathbf{x}^{\prime})\equiv&\left(\frac{3\mathbf{\hat{r}}\mathbf{\hat{r}}-\bm{\mathcal{I}}}{4\pi r^3}\right)_{\bar{V}_{\varepsilon}}\,,
\label{Dipole-Operator}
\end{align}
\end{subequations}
where $\bm{\mathcal{L}}_{\varepsilon}$ and $\bm{\mathcal{T}}_{\varepsilon}$ in the nabla forms are the longitudinal and transverse projection operators, respectively, and $\bm{\mathcal{O}}_{\varepsilon}$ is the dipole distribution. The projection decomposition of the current is a typical trick for deriving the potential equations in the Coulomb gauge \cite{Jackson1,Greiner}. Unfortunately, the distributional forms (integral kernels) of the projections have not been previously identified, making straightforward calculation on field projections difficult. Brill and Goodman first utilized the orthogonal projection to analyze the vector potential in the Coulomb gauge \cite{Brill}. However, the unsuitable spherical regularization did not contribute to their work. Surprisingly, the last distribution representation in Eq.~(\ref{Transverse-Operator}) is identical to Eq.~(\ref{Effective-Dipole-distribution}), which indicates that the normal discal regularization has a special meaning in field decomposition.

In the Coulomb or transverse gauge condition $\nabla\cdot$$\mathbf{A}$=0, the potential equations are \cite{Jackson1}(Sec. 6.3)
\begin{subequations}\label{Potential-Equations}
\begin{align}\label{Scalar-Equation}
\nabla^2\varphi_{\mathrm{C}}=&-\rho/\epsilon_0\,,\\
\label{Vector-Equation}
\Box\mathbf{A}_{\mathrm{C}}=&-\mu_0\mathbf{J}_{\perp}\,,
\end{align}
\end{subequations}
where $\Box$$\equiv$$\nabla^2$$-$$\partial^2/\partial (ct)^2$ is the wave operator, $\varphi_{\mathrm{C}}$ and $\mathbf{A}_{\mathrm{C}}$ are the scalar and vector potentials in the Coulomb gauge, respectively. The source term $\mathbf{J}_{\perp}$ in Eq.~(\ref{Vector-Equation}) is the transverse current density transformed from the current density $\mathbf{J}$ by the action of $\bm{\mathcal{T}}_{\varepsilon}$, i.e.,
\begin{align}
\mathbf{J}_{\perp}\equiv&\int_{\bar{V}_{\varepsilon}}\!\bm{\mathcal{T}}_{\varepsilon}(\mathbf{x},\mathbf{x}^{\prime})\cdot\mathbf{J}(\mathbf{x}^{\prime})\mathrm{d}^3x^{\prime}\nonumber\\
=&\,\bm{\mathcal{E}}\cdot\mathbf{J}+\int_{\bar{V}_{\varepsilon}}\!\bm{\mathcal{O}}_{\varepsilon}(\mathbf{x},\mathbf{x}^{\prime})\cdot\mathbf{J}(\mathbf{x}^{\prime})\mathrm{d}^3x^{\prime}\,.
\label{Transverse-Nonlocal-DC}
\end{align}
It can be checked that $\mathbf{J}_{\perp}$ has the same form as the static external displacement current given in Eq.~(\ref{General-Continous-Nonlocal-DC-2}). The solution to Eq.~(\ref{Vector-Equation}) in all space $V_{\infty}$ can be derived using a retarded Green's function,
\begin{align}
G(\mathbf{x},t;\mathbf{x}^{\prime},t^{\prime})=\frac{1}{|\mathbf{x}-\mathbf{x}^{\prime}|}\delta(t-t^{\prime}-|\mathbf{x}-\mathbf{x}^{\prime}|/c)\,.
\label{Green-Function}
\end{align}
In this way, the vector potential can be expressed as
\begin{align}
\mathbf{A}_{\mathrm{C}}
=&\frac{\mu_0}{4\pi}\int_{V_{\infty}}\!G(\mathbf{x},t;\mathbf{x}^{\prime},t^{\prime})\mathbf{J}_{\perp}(\mathbf{x}^{\prime},t^{\prime})\mathrm{d}^3x^{\prime}\mathrm{d}t^{\prime}\nonumber\\
=&\frac{\mu_0}{4\pi}\int_{V_{\infty}}\!\mathrm{d}^3x^{\prime}\!\int_{-\infty}^{\infty}\mathrm{d}t^{\prime}G(\mathbf{x},t;\mathbf{x}^{\prime},t^{\prime})\nonumber\\
\times&\int_{\bar{V}^{\prime}_{\varepsilon}}\!\bm{\mathcal{T}}_{\varepsilon}(\mathbf{x}^{\prime},\mathbf{x}^{\prime\prime})\cdot\mathbf{J}(\mathbf{x}^{\prime\prime},t^{\prime})
\mathrm{d}^3x^{\prime\prime}\nonumber\\
=&\frac{\mu_0}{4\pi}\int_{V_{\infty}}\!\frac{\mathbf{J}_{\perp}(\mathbf{x}^{\prime},t_r)}{r}\mathrm{d}^3x^{\prime}\,,
\label{Vector-Potential-Coulomb-1}
\end{align}
where $t_r$=$t$$-$$r/c$. Equation (\ref{Vector-Potential-Coulomb-1}) infers that the vector potential formally arises from the retarded static external displacement current. By insertion of Eq.~(\ref{Transverse-Nonlocal-DC}), the vector potential can be expressed as
\begin{align}
\mathbf{A}_{\mathrm{C}}=&\frac{\mu_0}{4\pi}\int_{V_{\infty}}\!\frac{\bm{\mathcal{E}}\cdot\mathbf{J}(\mathbf{x}^{\prime},t_r)}{r}\mathrm{d}^3x^{\prime}\nonumber\\
+&\frac{\mu_0}{4\pi}\int_{V_{\infty}}\mathrm{d}^3x^{\prime\prime}\!\int_{\bar{V}^{\prime\prime}_{\varepsilon}}\frac{3\mathbf{\hat{r}}^{\prime}\mathbf{\hat{r}}^{\prime}-\bm{\mathcal{I}}}{4\pi rr^{\prime3}}\cdot\mathbf{J}\left(\mathbf{x}^{\prime\prime},t_r\right)\mathrm{d}^3x^{\prime}\,,
\label{Vector-Potential-Coulomb-2}
\end{align}
where $r^{\prime}$=$|\mathbf{x}^{\prime}$$-$$\mathbf{x}^{\prime\prime}|$, and $R$=$|\mathbf{x}$$-$$\mathbf{x}^{\prime\prime}|$ as before. The region $\bar{V}^{\prime\prime}_{\varepsilon}$, the domain of the dipole distribution, is equivalently mapped into the integral region.  For the subsequent calculation, we adopt a tilted-disk regularization with the disk's normal along a specific radial direction; the detailed steps are given in Appendix C. A final simplification gives rise to
\begin{align}
\mathbf{A}_{\mathrm{C}}=&\mu_0\int_{V_{\infty}}\!\frac{\mathbf{J}(\mathbf{x}^{\prime},t_r)}{4\pi r}\mathrm{d}^3x^{\prime}-\mu_0\int_{V_{\infty}}\!\frac{\mathbf{J}(\mathbf{x}^{\prime},t_r)\cdot\mathbf{\hat{r}}\mathbf{\hat{r}}}{4\pi r}\mathrm{d}^3x^{\prime}\nonumber\\
+&\int_{V_{\infty}}\frac{\mathrm{d}^3x^{\prime}(3\mathbf{\hat{r}}\mathbf{\hat{r}}-\bm{\mathcal{I}})}{4\pi\epsilon_0 r^3}\cdot\!\int_{t_r}^t\mathbf{J}(\mathbf{x}^{\prime},\tau)(t-\tau)\mathrm{d}\tau\,.
\label{Vector-Potential-Coulomb-3}
\end{align}

Equation~(\ref{Vector-Potential-Coulomb-3}) was derived by Jackson via a gauge transformation of the Lorenz gauge \cite{Jackson4}, and it also serves as a verification of the transverse distribution here.
\subsection{Expression of instantaneous displacement current}
The displacement current in this case can be derived from $\mathbf{E}=-\nabla\varphi_{\mathrm{C}}-\dot{\mathbf{A}}_{\mathrm{C}}$ as
\begin{align}
\mathbf{J}_{\mathrm{D}}=-\epsilon_0(\nabla\dot{\varphi}_{\mathrm{C}}+\ddot{\mathbf{A}}_{\mathrm{C}})\,.
\label{General-Total-DC}
\end{align}
The first term on the rhs. of Eq.~(\ref{General-Total-DC}) represents the instantaneous scalar potential and is exactly the longitudinal component of the negative current \cite{Jackson1} (Sec. 6.3), i.e.,
\begin{align}
\epsilon_0\nabla\dot{\varphi}_{\mathrm{C}}=&\,\mathbf{J}_{\parallel}
=\int_{V_{\infty}}\!\bm{\mathcal{L}}_{\varepsilon}(\mathbf{x},\mathbf{x}^{\prime})\cdot\mathbf{J}(\mathbf{x}^{\prime})\mathrm{d}^3x^{\prime}\nonumber\\
=&\,\mathbf{J}-\mathbf{J}_{\perp}\,,
\label{Longitudinal-Current}
\end{align}
where the third identity in Eq.~(\ref{Transverse-Operator}) is used. In addition, by inserting the identity
\begin{align}
\frac{\partial^2}{\partial t^2}&\left[\int_{t_r}^t\mathbf{J}(\mathbf{x}^{\prime},\tau)(t-\tau)\mathrm{d}\tau\right]\nonumber\\
=&\,\mathbf{J}(\mathbf{x}^{\prime},t)-\mathbf{J}(\mathbf{x}^{\prime},t_r)-\dot{\mathbf{J}}(\mathbf{x}^{\prime},t_r)r/c\,,
\label{Second-Time-Derivative}
\end{align}
one can transform the displacement current in Eq.~(\ref{General-Total-DC}) into
\begin{subequations}\label{R-DC-General-Distribution}
\begin{align}\label{R-DC-General-Distribution-1}
\mathbf{J}^{\mathrm{int}}_{\mathrm{D}}=&-\mathbf{J}\,,\\
\mathbf{J}^{\mathrm{ext}}_{\mathrm{D}}=&\int_{V_{\infty}}\left[\frac{\dot{\mathbf{J}}^{\prime}_{\mathrm{ret}}\cdot(3\mathbf{\hat{r}}\mathbf{\hat{r}}-\bm{\mathcal{I}})}{4\pi cr^2}
+\frac{\mathbf{\hat{r}}\times(\mathbf{\hat{r}}\times\ddot{\mathbf{J}}^{\prime}_{\mathrm{ret}})}{4\pi c^2r}\right]\mathrm{d}^3x^{\prime}\nonumber\\
+&\,\bm{\mathcal{E}}\cdot\mathbf{J}+\int_{\bar{V}_{\varepsilon}}\frac{\mathbf{J}^{\prime}_{\mathrm{ret}}}
{4\pi r^3}\cdot(3\mathbf{\hat{r}}\mathbf{\hat{r}}-\bm{\mathcal{I}})\mathrm{d}^3x^{\prime}\,,
\label{R-DC-General-Distribution-2}
\end{align}
\end{subequations}
where $\mathbf{J}^{\prime}_{\mathrm{ret}}$ is an abbreviation for $\mathbf{J}(\mathbf{x}^{\prime},t_r)$, another convention consistently used throughout this paper. The displacement current as a whole is similar to that from Heras \cite{Heras2} when we choose the spherical regularization by setting $\bm{\mathcal{E}}$=$2\bm{\mathcal{I}}/3$. The first term in the second line of Eq.~(\ref{R-DC-General-Distribution-2}) is the localized component. 
Moreover, the wave equation for the external displacement current is
\begin{align}
\label{Vector-Displacement-Equation}
\Box\mathbf{J}^{\mathrm{ext}}_{\mathrm{D}}=-\nabla\times\nabla\times\mathbf{J}\,,
\end{align}
which can be directly derived from the wave equation for the magnetic field \cite{Jackson1}(Sec. 6.5),
$\Box\mathbf{H}=-\nabla\times\mathbf{J}$.
By applying Green's function (\ref{Green-Function}), a differential solution can be found to be
\begin{align}
\label{Vector-Displacement-Integration}
\mathbf{J}^{\mathrm{ext}}_{\mathrm{D}}=\,&\nabla\times\nabla\times\int_{V_{\infty}}\frac{\mathbf{J}^{\prime}_{\mathrm{ret}}}{4\pi r}\mathrm{d}^3x^{\prime}\,,
\end{align}
which is quite understandable since the integral is the vector potential in the Lorentz gauge. The equality between Eqs.~(\ref{R-DC-General-Distribution-2}) and (\ref{Vector-Displacement-Integration}) can be verified in the spherical regularization, as is omitted here.

\section{Interpretation to the Biot-Savart law}
\subsection{Instantaneous Biot-Savart integral}
Apart from Eq.~(\ref{Vector-Potential-Coulomb-1}), the vector potential in the Coulomb gauge can also appear as another form \cite{Maxwell,Weber,McDonald3}
\begin{align}
\label{New-Coulomb-Potential}
\mathbf{A}_{\mathrm{C}^{\prime}}=&\,\frac{\mu_0}{4\pi}\int_{V_{\infty}}\frac{\mathbf{J}^{\mathrm{ext}}_{\mathrm{D}}(\mathbf{x}^{\prime})}{r}\mathrm{d}^3x^{\prime}\,,
\end{align}
where $\mathbf{A}_{\mathrm{C}^{\prime}}$ actually is equivalent to $\mathbf{A}_{\mathrm{C}}$. Equation (\ref{New-Coulomb-Potential}) requires that the magnetic field decays faster than $1/r$ at infinity, which is a standard condition in electromagnetism. The vorticity of this vector potential directly yields a well-known result
\begin{align}
\label{New-H}
\mathbf{H}=\int_{V_{\infty}}\frac{\mathbf{J}^{\mathrm{ext}}_{\mathrm{D}}(\mathbf{x}^{\prime})\times\mathbf{\hat{r}}}{4\pi r^2}\mathrm{d}^3x^{\prime}\,,
\end{align}
which can be taken as an instantaneous Biot-Savart integral, an integral form of the Amp\`{e}re-Maxwell law. In fact, the form of Eq.~(\ref{New-H}) was long regarded as having little practical significance \cite{Griffiths-Heald} and, as a result, does not appear in most textbooks. A contemporary analysis can be found in Ref. \cite{McDonald3}. Inserting Eq.~(\ref{Vector-Displacement-Integration}) into Eq.~(\ref{New-H}) leads to a double-curl form, i.e.,
\begin{subequations}\label{New-H-1}
\begin{align}\label{New-H-11}
\mathbf{H}=\,&\nabla\times\int_{V_{\infty}}\frac{\nabla^{\prime}\times\mathbf{H}_0(\mathbf{x}^{\prime})}{4\pi r}\mathrm{d}^3x^{\prime}\,,\\
\mathbf{H}_0=\,&\nabla\times\int_{V_{\infty}}\frac{\mathbf{J}(\mathbf{x}^{\prime},t_r)}{4\pi r}\mathrm{d}^3x^{\prime}\,.
\label{New-H-12}
\end{align}
\end{subequations}
Since the fields involved can also be considered to vanish faster than $1/r$ in space, the rotational component of Helmholtz decomposition in Eq.~(\ref{New-H-11}) is unique \cite{Griffiths}. Therefore, $\mathbf{H}$ is the same as $\mathbf{H}_0$, and it becomes
\begin{align}
\mathbf{H}=\mathbf{H}_0=&\frac{1}{4\pi}\!\int_{V_{\infty}}\!\nabla\times\frac{\mathbf{J}(\mathbf{x}^{\prime},t_r)}{r}\mathrm{d}^3x^{\prime}\nonumber\\
=&\frac{1}{4\pi}\!\int_{V_{\infty}}\!\left(\frac{\mathbf{J}^{\prime}_{\mathrm{ret}}}{r^2}+\frac{\dot{\mathbf{J}}^{\prime}_{\mathrm{ret}}}{cr}\right)\times\mathbf{\hat{r}}\mathrm{d}^3x^{\prime}\,,
\label{J-H-Equation}
\end{align}
after performing the curl operation in Eq.~(\ref{New-H-12}). This is Jefimenko's equation, a generalized form of the Biot-Savart law. One may think that the indirect derivation of Eq.~(\ref{J-H-Equation}) does not account for the role of the displacement current. In fact, the integral representation of the external displacement current (\ref{R-DC-General-Distribution-2}) can also be verified directly when applied to Eq.~(\ref{New-H}). A subsequent simplification at the end of Appendix C, recovers Jefimenko's equation (\ref{J-H-Equation}) for the magnetic field. 

The instantaneous Biot-Savart integral does not explicitly involve the current, but it reveals a spatial tranformation between fields. The physical meaning of Eq.~(\ref{New-H}) is subtle in consideration of Maxwell's current equivalence \cite{Heras1}, and will be revisited in the following subsection.

\subsection{Current equivalence under the quasi-steady-state and quasi-static-state conditions}
Substitution Eq. (\ref{R-DC-General-Distribution-1}) for the rhs. of Eq. (\ref{A-M-Law}) yields
\begin{align}
\nabla\times\mathbf{H}=\mathbf{J}_{\mathrm{tot}}=\mathbf{J}^{\mathrm{ext}}_{\mathrm{D}}\,,
\label{A-M-Law-New}
\end{align}
where $\mathbf{J}_{\mathrm{tot}}$ is the total current as before \cite{Huang1} and will be used in parallel with the external displacement current in the following context. The rhs. of Eq.~(\ref{A-M-Law-New}) does not include the current; however, the original form of the total current does. The discrepancy in the representative form could be one of the underlying sources of the debate over the displacement current. Furthermore, the formal disappearance of the current in Maxwell's equations implies its indirect connection to the magnetic field. It is tantamount to stating that the external displacement current or the total current is possibly the only (field) source of magnetic field \cite{Huang2}!

In the following discussion, two weakly time-dependent cases will be considered: the quasi-steady state and the quasi-static state, both of which are redefined here. However, the associated current and charge conditions are subtle \cite{Griffiths-Heald,Larsson}. For clarity, only constraints on the current are adopted here. Firstly, a quasi-steady state is defined by $\ddot{\mathbf{J}}$=0, which implies $\ddot{\mathbf{H}}$=0 from the Biot-Savart law below. This condition supports the steady charging and discharging process of a plate capacitor and infers
\begin{align}
\mathbf{J}(\mathbf{x}^{\prime},t)=\mathbf{J}(\mathbf{x}^{\prime},t_r+r/c)=\mathbf{J}(\mathbf{x}^{\prime},t_r)+\dot{\mathbf{J}}(\mathbf{x}^{\prime},t_r)r/c\,,
\label{Steady-Condition}
\end{align}
which leads to
\begin{align}
\mathbf{J}_{\mathrm{tot}}=\mathbf{J}^{\mathrm{ext}}_{\mathrm{D}}=\mathbf{J}_{\perp}=\mathbf{J}-\mathbf{J}_{\parallel}\,,\,\,\,\,\mathbf{J}_{\mathrm{D}}=-\mathbf{J}_{\parallel}\,,
\label{QSteady-TotalD-Decomposition}
\end{align}
from Eqs.~(\ref{Longitudinal-Current}) and (\ref{R-DC-General-Distribution}). In this case, Eq.~(\ref{General-Continous-Nonlocal-DC-2}) is valid and thus allows the expression in Eq.~(\ref{Appc-New-H-Expansion}) to be simplified into
\begin{align}
\mathbf{H}_{\mathrm{qs}}=&\int_{V_{\infty}}\!\frac{[\bm{\mathcal{E}}\cdot\mathbf{J}(\mathbf{x}^{\prime\prime},t)]\times\mathbf{\hat{R}}}{4\pi R^2}\mathrm{d}^3x^{\prime\prime}\nonumber\\
+&\int_{V_{\infty}}\frac{\mathrm{d}^3x^{\prime\prime}}{4\pi}\!\int_{\bar{V}^{\prime\prime}_{\varepsilon}}\frac{(3\mathbf{\hat{r}}^{\prime}\mathbf{\hat{r}}^{\prime}
-\bm{\mathcal{I}})\cdot\mathbf{J}\left(\mathbf{x}^{\prime\prime},t\right)}{4\pi r^2r^{\prime3}}\times\mathbf{\hat{r}}\mathrm{d}^3x^{\prime}\,.
\label{QSteady-Magnetic-Field}
\end{align}
The final result of the integral in the second line vanishes according to the calculation in Appendix C. Equation~(\ref{QSteady-Magnetic-Field}) then becomes
\begin{align}
\mathbf{H}_{\mathrm{qs}}=&\int_{V_{\infty}}\!\frac{[(\bm{\mathcal{I}}-\hat{\mathbf{R}}\hat{\mathbf{R}})\cdot\mathbf{J}(\mathbf{x}^{\prime\prime},t)]\times\mathbf{\hat{R}}}{4\pi R^2}\mathrm{d}^3x^{\prime\prime}\nonumber\\
=&\int_{V_{\infty}}\!\frac{\mathbf{J}(\mathbf{x}^{\prime},t)\times\mathbf{\hat{r}}}{4\pi r^2}\mathrm{d}^3x^{\prime}\,(\mathbf{x}^{\prime\prime}\rightarrow\mathbf{x}^{\prime})\,,
\label{QSteady-Magnetic-Field-BS}
\end{align}
which is the classical Biot-Savart law, as confirmed by other methods \cite{Zhu,Weber,Griffiths-Heald}. Although it can be directly obtained by combining Eqs.~(\ref{J-H-Equation}) and (\ref{Steady-Condition}), the above derivation reveals how the current replaces the localized displacement current in expressing the magnetic field. 
Since $\mathbf{J}_{\mathrm{D}}$=$-$$\mathbf{J}_{\parallel}$ and the longitudinal component of the current as a whole does not cause a magnetic field due to Eq.~(\ref{Vector-Equation}), a comparison between Eqs.~(\ref{New-H}) and (\ref{QSteady-Magnetic-Field-BS}) clearly demonstrates the equivalence between the current and the external displacement current. It should be noted that the equivalence relation is now refined according to the source roles of the two types of currents. Thus, Cullwick's paradox can be effectively explained by substituting the current for the (external) displacement current, rather than combining them. The equivalence simplifies the physical law and directly expresses it in terms of measurable quantity, i.e., the current; this simplification, nevertheless, hides the intrinsic connection between fields. 

The quasi-static state here refers to a special magnetoquasistatics model \cite{Larsson}, characterized by at most a linear time variation of the current, which necessitates an additional solenoidal condition, $\nabla\cdot\mathbf{J}$=0, in the quasi-steady state. In this case, the first equality in Eq.~(\ref{Longitudinal-Current}) gives rise to $\mathbf{J}_{\mathrm{D}}$=$-$$\mathbf{J}_{\parallel}$=0 because $\dot{\rho}$$=$0$\Rightarrow$$\dot{\varphi}_{\mathrm{C}}$=0. It conveys a surprising result from Eq.~(\ref{DC-General-Case}): $\mathbf{J}^{\mathrm{ext}}_{\mathrm{D}}$=$\mathbf{J}$, causing the external displacement current to have a completely localized distribution that is formally indistinguishable from the real current. Paradoxically, it is commonly believed that the relation $\mathbf{J}_{\mathrm{D}}$=0 implies that the real current, not the external displacement current field, plays a role. The localization of the nonlocal external displacement current makes a perplexing coincidence. This result, as presented in Appendix D, can be alternatively derived from Eq.~(\ref{General-Continous-Nonlocal-DC-2}) or Eq.~(\ref{Transverse-Nonlocal-DC}). Equation (\ref{Appd-Final-T0}) indicates that nearly all nonlocal components of the external displacement current cancel each other out, leaving only a local contribution from an infinitesimal, needle-like neighborhood along the current path. 

We may question the peculiar result $\mathbf{J}_{\mathrm{D}}$=0, after all, its internal and external components are quite different. This result is actually associated with the statistical definitions of macroscopic quantities over their microscopic counterparts that do not satisfy the continuum approximation. Macroscopic quantities can be taken as a spatial average of those in the microscopic world at the level of electrons and nuclei \cite{Jackson1} (Sec. 6.6), such as the current $\mathbf{J}$$\equiv$$\langle\mathbf{j}\rangle$ (also equal to $-\langle\mathbf{j}_{\mathrm{D}}^{\mathrm{int}}\rangle$), where the lowercase letters are used for the microscopic objects. The total current has a form $\mathbf{J}_{\mathrm{tot}}$$\equiv$$\langle\mathbf{j}+\mathbf{j}_{\mathrm{D}}^{\mathrm{int}}+\mathbf{j}_{\mathrm{D}}^{\mathrm{ext}}\rangle$=$\langle\mathbf{j}_{\mathrm{D}}^{\mathrm{ext}}\rangle$=$\mathbf{J}_{\mathrm{D}}^{\mathrm{ext}}$, where the identity $\mathbf{j}$$+$$\mathbf{j}_{\mathrm{D}}^{\mathrm{int}}$=0 strictly holds. However, $\mathbf{j}_{\mathrm{D}}^{\mathrm{int}}$$+$$\mathbf{j}_{\mathrm{D}}^{\mathrm{ext}}$$\neq$0 at the microscopic level of at least the scale of the mean free path of free charges, but they are approximately zero at the macroscopic scale: $\langle\mathbf{j}_{\mathrm{D}}^{\mathrm{int}}$$+$$\mathbf{j}_{\mathrm{D}}^{\mathrm{ext}}\rangle$$\approx$0 in conductors. Therefore, the matter source $\mathbf{J}$$\approx$$\mathbf{J}_{\mathrm{tot}}$, not the field source $\mathbf{J}_{\mathrm{D}}^{\mathrm{ext}}$$=$$\mathbf{J}_{\mathrm{tot}}$, is intuitively taken for the vorticity of the magnetic field. Therefore, the problem of the plate capacitor and Griffiths' question can be solved, because the external displacement current does exist inside any Amperian loop around a wire, where one might think there is nothing but the real current. On the other hand, evidently, the total current forms a closed vortex field throughout all space, including within the conductors. Maxwell's assertion regarding the closure issue of the total current is fundamentally sound.

The last question is whether the current equivalence holds in a general case beyond the quasi-steady state. Interestingly, the curl of vector potential in Eq.~(\ref{Vector-Potential-Coulomb-1}) gives rise to
\begin{align}
\mathbf{H}=\frac{1}{4\pi}\!\int_{V_{\infty}}\!\left(\frac{\mathbf{J}^{\prime}_{\bot\mathrm{ret}}}{r^2}+\frac{\dot{\mathbf{J}}^{\prime}_{\bot\mathrm{ret}}}{cr}\right)\times\mathbf{\hat{r}}\mathrm{d}^3x^{\prime}\,,
\label{SDC-H-Equation}
\end{align}
where $\mathbf{J}^{\prime}_{\bot\mathrm{ret}}$=$\mathbf{J}_{\bot}(\mathbf{x}^{\prime},t_r)$. This is expected, since only the transverse (vortex) component of the current contributes to the magnetic field. Comparing this expression to Jefimenko's equation (\ref{J-H-Equation}), we can also state that the current is equivalent to a static external displacement current (an unreal displacement current) and becomes equal to it in a quasi-static state.

\section{Discussions}
The idealized concept of a point charge leads to the peculiar point singularity in the internal displacement current. The point-like 0-D treatment of the charge is not the only way to decompose the displacement current. One possible method to avoid the singularity is to model the point as a smooth 3-D distribution, such as a spherically symmetric Gaussian function. In this case, both the internal displacement current and the current are continuously distributed. However, such modeling is not essential, as it neither affects the decomposition of the displacement current nor disrupts the cancellation of the current. Subtracting the internal singularity, the displacement current formally becomes the electric counterpart to the time derivative of magnetic field in Maxwell's equations, acting as a vortex-field source. This further clarifies the mutual induction between vortex electric and vortex magnetic fields. 

Although the displacement current may disappear in potential equations under other gauge conditions, its role in bridging the current and magnetism remains unaffected. On this point, although it is seldom used in calculations, the instantaneous Biot-Savart integral is essential for demonstrating the current equivalence and the causal relationship. In the conventional framework for interpreting electromagnetic induction, the total current can be considered the direct causal source of the magnetic field, according to the aforementioned definition of ``source." However, this cause-and-effect viewpoint cannot address the problem of simultaneity, an issue inherent in the conventional interpretation. This issue will be further discussed in Appendix E, where another worldview will be introduced to understand Maxwell's equations. 

\section{Conclusions}
A discal regularization for the typical total dipole distribution is proposed as an appropriate generalization scheme to characterize the displacement current. This regularization also formulates the projection operators as distributions, which can be employed to obtain their integral representations of projections, as has been done for the Coulomb vector potential and displacement current. Importantly, the displacement current can be decomposed into an external field and an internal singular component with an arbitrary regularization. The internal distribution counteracts the real current, so it simplifies the Amp\`{e}re-Maxwell law, and formally recasts the magnetic field as an instantaneous Biot-Savart integral of the remaining external distribution. By either the simplified Amp\`{e}re-Maxwell law or its integral form, the external displacement current can reasonably be considered the sole local cause of the magnetic field generation. The current equivalence under the steady state, a substitution relation, is a result of a simplified integral of the external displacement current. Especially under the quasi-static condition, one may easily misjudge the magnetic mechanism in that the external displacement current is exactly equal to the electric current. The field-based reinterpretation facilitates a settlement of several disputes between the two currents, albeit at the expense of challenging the widely accepted notion that electric current directly generates magnetic fields. 

\appendix
\section{Disk distribution}
Equation~(\ref{Local-DC-Limit-2}) provides a sequence of ordinary functions of $\mathbf{x}$, parameterized by $r^{\prime}$, which converges to a distribution regularized in an infinitesimal discal zone. This can be simply verified by its integration in the limit $r^{\prime}\rightarrow 0$. By means of the volume element $\pi\mathrm{d}r_{\parallel}\mathrm{d}r^2_{\perp}$ in the cylindrical coordinate system, the integration of the first part in Eq.~(\ref{Local-DC-Limit-2}) over $V^{\prime}_{\mathrm{d}}$, referring to Fig.~\ref{fig-2}, gives rise to
\begin{align}
\lim_{r^{\prime}\rightarrow 0}&\int_{-\infty}^{\infty}\frac{r_{\parallel}\mathrm{d}r_{\parallel}}{r^{\prime}}\mathrm{rect}\left(\frac{r_{\parallel}}{r^{\prime}}\right)\int_0^{r^2_{\mathrm{d}}}\frac{\pi \mathrm{d}r^2_{\perp}}{2\pi(r^2_{\perp}+r_{\parallel}^2)^{\frac{3}{2}}}\nonumber\\
=&1-\lim_{r^{\prime}\rightarrow 0}\int_0^{r^{\prime}}\frac{r_{\parallel}\mathrm{d}r_{\parallel}}{r^{\prime}(r^2_{\mathrm{d}}+r^2_{\parallel})^{1/2}}=1\,.
\label{Appa-Dirac-Integration1}
\end{align}
Hence, the integral function in Eq. (\ref{Appa-Dirac-Integration1}) is a 3-D $\delta$ distribution $\delta(\mathbf{r})$. Similarly, replacing $r_{\parallel}$ by $R_{\parallel}$ leads to the same result. 

Likewise, it can be seen that the second line of Eq. (\ref{Relativity-Difference-Q-D}) evolves into the same $\delta$ function by a substitution $\gamma r_{\parallel}$$\rightarrow$$r_{\parallel}$, which can simply be verified by
\begin{align}
\frac{1}{2\pi}&\int_0^{r^2_{\mathrm{d}}}\frac{\pi \gamma r_{\parallel}\mathrm{d}(r_{\perp})^2}{r_{\mathrm{r}}^3}=1-\frac{\gamma r_{\parallel}}{\sqrt{r^2_{\mathrm{d}}+\gamma^2 r^2_{\parallel}}}\,.
\label{Appa-Dirac-Integration2}
\end{align}
A further verification requires multiplying the sequence by a smooth test function and taking the limit after integration, the details of which are omitted here for simplicity. 

\section{Regularization of the dipole distribution in the difference-quotient method}
In this appendix, the difference-quotient method of regularization to the dipole distribution will be presented and applied in some common cases. According to the treatment in Sec. II of the displacement current for a moving point charge, the difference-quotient method can be summarized in four steps: 1. considering the difference quotient of the dipole distribution in the $n$th coordinate direction ($n$=1, 2, 3) between two points $\mathbf{x}^{\prime\prime}_n$ and $\mathbf{x}^{\prime}_n$, with $\mathbf{r}^{\prime}_n$=$\mathbf{x}^{\prime\prime}_n$$-$$\mathbf{x}^{\prime}_n$, $\mathbf{r}_n$=$\mathbf{x}$$-$$\mathbf{x}^{\prime}_n$; 2. symmetrically delineating a specified internal region $V^{\prime}_{\varepsilon}$ clamped between the two displaced points;
3. performing a limit operation $r_n^{\prime}$$\rightarrow$0 after integrating the difference quotient over the specified region; 4. repeating the steps above for the other two orthogonal directions to perform a complete vector decomposition of the tensor-valued distribution. When these computations are completed, the internal dipole distribution will be represented by nine coefficients in matrix form.

Although regularization can be characterized by an arbitrary $V^{\prime}_{\varepsilon}$, practical applications typically involve a limited set of symmetrical convex geometries. It is therefore reasonable to require that they exhibit at least some basic mirror symmetries along three orthogonal directions, meaning they belong to or include the point group $mmm$ ($D_{2h}$). Under this symmetry, only the longitudinal components of the difference quotient remain, and the contribution from the term $\mathbf{r}_{n\parallel}$ is equal to that of $\mathbf{R}_{n\parallel}$. So, the effective difference quotient can be simplified as $-2\mathbf{r}_{n\parallel}/r_n^3$ divided by the characteristic length $r^{\prime}_n$ according to Eq.~(\ref{Local-DC-Limit-2}). Let the principal axes of the geometry be along the intersecting lines of the three symmetrical planes as $x_{\mathrm{p}i}$ with their orthogonal unit vectors $\mathbf{\hat{e}}_{\mathrm{p}i}$ ($i$=1, 2, 3), the singular distribution can then be formulated by a characteristic tensor $\bm{\mathcal{D}}_{\varepsilon}$ as
\begin{align}
\left(\nabla\nabla\frac{1}{r}\right)_{\!\!V^{\prime}_{\varepsilon}}=&-\sum_{n=1}^3\mathbf{\hat{e}}_{\mathrm{p}n}\mathbf{\hat{e}}_{\mathrm{p}n}\lim_{r_n^{\prime}\rightarrow0}
\left(\frac{2r_{n\parallel}}{r^{\prime}_nr_n^3}\right)_{\!\!V^{\prime}_{\varepsilon}}\nonumber\\
\equiv&-4\pi\delta(\mathbf{r})\sum_{n=1}^3\mathcal{D}_{\varepsilon nn}\mathbf{\hat{e}}_{\mathrm{p}n}\mathbf{\hat{e}}_{\mathrm{p}n}\nonumber\\
=&-4\pi\delta(\mathbf{r})\bm{\mathcal{D}}_{\varepsilon}\,,
\label{Internal-Regularization}
\end{align}
where $r_{n\parallel}$=$\mathbf{\hat{e}}_{\mathrm{p}n}\cdot\mathbf{r}_n$. The representative matrix of $\bm{\mathcal{D}}_{\varepsilon}$ is thus diagonalized, with
\begin{align}
&\mathcal{D}_{\varepsilon nn}=\int_{V^{\prime}_{\varepsilon}}\lim_{r^{\prime}_n\rightarrow 0}\frac{r_{n\parallel}\mathrm{d}^3x}{2\pi r^{\prime}_nr^3_n}
=\lim_{r^{\prime}_n\rightarrow 0}\int_{V^{\prime}_{\varepsilon}}\frac{r_{n\parallel}\mathrm{d}^3x}{2\pi r^{\prime}_nr^3_n}
\nonumber\\
&=\int_{V^{\prime}_{\varepsilon}(1)}\frac{\mathbf{\hat{e}}_{\mathrm{p}n}\cdot\mathbf{\bar{r}}_n}{2\pi\bar{r}_n^3}\mathrm{d}^3\bar{x}
=\mathbf{\hat{e}}_{\mathrm{p}n}\cdot\int_{V^{\prime}_{\varepsilon}(1)}-2\epsilon_0\frac{-\mathbf{\bar{r}}_n\mathrm{d}^3\bar{x}}{4\pi\epsilon_0\bar{r}_n^3}
\,,
\label{Dnn}
\end{align}
where the scaled variables with bars are defined by $\mathbf{\bar{r}}_n$=$\mathbf{r}_n/r^{\prime}_n$, $\mathbf{\bar{x}}$=$\mathbf{x}/r^{\prime}_n$, $r^{\prime}_n$ represents the characteristic length scale of $V^{\prime}_{\varepsilon}$ in the $\mathbf{\hat{e}}_{\mathrm{p}n}$ direction, and $V^{\prime}_{\varepsilon}(1)$ is $V^{\prime}_{\varepsilon}$ at $r^{\prime}_n$=1. Equation~(\ref{Dnn}) shows that the diagonal elements can be taken as the electric fields at the edge points $\mathbf{x}^{\prime}_n$ exerted by the uniformly charged ($\rho$=$-2\epsilon_0$) body $V^{\prime}_{\varepsilon}$ with a unit characteristic length of $r^{\prime}_n$=1, This implies that $\mathcal{D}_{\varepsilon nn}$ are nonnegative parameters that depend solely on the shape of the specified geometry, rendering the limit $r^{\prime}_{n}$$\rightarrow$0 in Eq.~(\ref{Dnn}) unnecessary.

For the spherical regularization, the spherical symmetry permits us only to consider a constant tensor $\bm{\mathcal{D}}_{\mathrm{s}}$ with the form $\bm{\mathcal{D}}_{\mathrm{s}}$=$\mathcal{D}_{\mathrm{s}}$$\bm{\mathcal{I}}$. Setting a ball $V^{\prime}_{\mathrm{s}}$ centered at $\mathbf{x}_0^{\prime}$=$(\mathbf{x}^{\prime}$$+$$\mathbf{x}^{\prime\prime})/2$, with a radius $r^{\prime}/2$, Eq.~(\ref{Dnn}) can be expressed in the spherical coordinate frame ($r_{\mathrm{p}}$, $\theta$, $\phi$) as
\begin{align}
&\mathcal{D}_{\mathrm{s}}=\frac{1}{2\pi r^{\prime}}\int_{V^{\prime}_{\mathrm{s}}}\frac{r_{\parallel}}{r^3}
\mathrm{d}^3x=\frac{1}{2\pi r^{\prime}}\int_{V^{\prime}_{\mathrm{s}}}\frac{r_{\parallel}}{r^3}
r_{\mathrm{p}}^2\sin\theta\mathrm{d}\theta\mathrm{d}\phi\mathrm{d}r_{\mathrm{p}}\nonumber\\
&=\frac{1}{r^{\prime}}\int_0^{r^{\prime}/2}r_{\mathrm{p}}^2\mathrm{d}r_{\mathrm{p}}\int_{-1}^{1}
\frac{\mathrm{d}\mathit{\Theta}(r^{\prime}/2-r_{\mathrm{p}}\mathit{\Theta})}{[(r^{\prime}/2)^2+r_{\mathrm{p}}^2-r_{\mathrm{p}}r^{\prime}\mathit{\Theta}]^{3/2}}
\label{Appb-S-Regularization-1}
\end{align}
with $\mathit{\Theta}$=$-\cos\theta$. By a radial scaling of $r_{\mathrm{p}}$ as $\xi$=$2r_{\mathrm{p}}/r^{\prime}$, above integral turns into
\begin{align}
\mathcal{D}_{\mathrm{s}}=\int_0^{1}\xi^2\mathrm{d}\xi\int_{-1}^{1}\frac{(1-\xi\mathit{\Theta})\mathrm{d}\mathit{\Theta}}{2(1+\xi^2-2\xi\mathit{\Theta})^{3/2}}
=\frac{1}{3}\,,
\label{Appb-S-Regularization-2}
\end{align}
where the inner integration over $\mathit{\Theta}$ is the number 1. Expression~(\ref{Appb-S-Regularization-2}) corresponds to the last term in Eq.~(\ref{Ball-Regularization}).
\begin{figure}[ht]
\centering
\begin{overpic}
[scale=1.0]{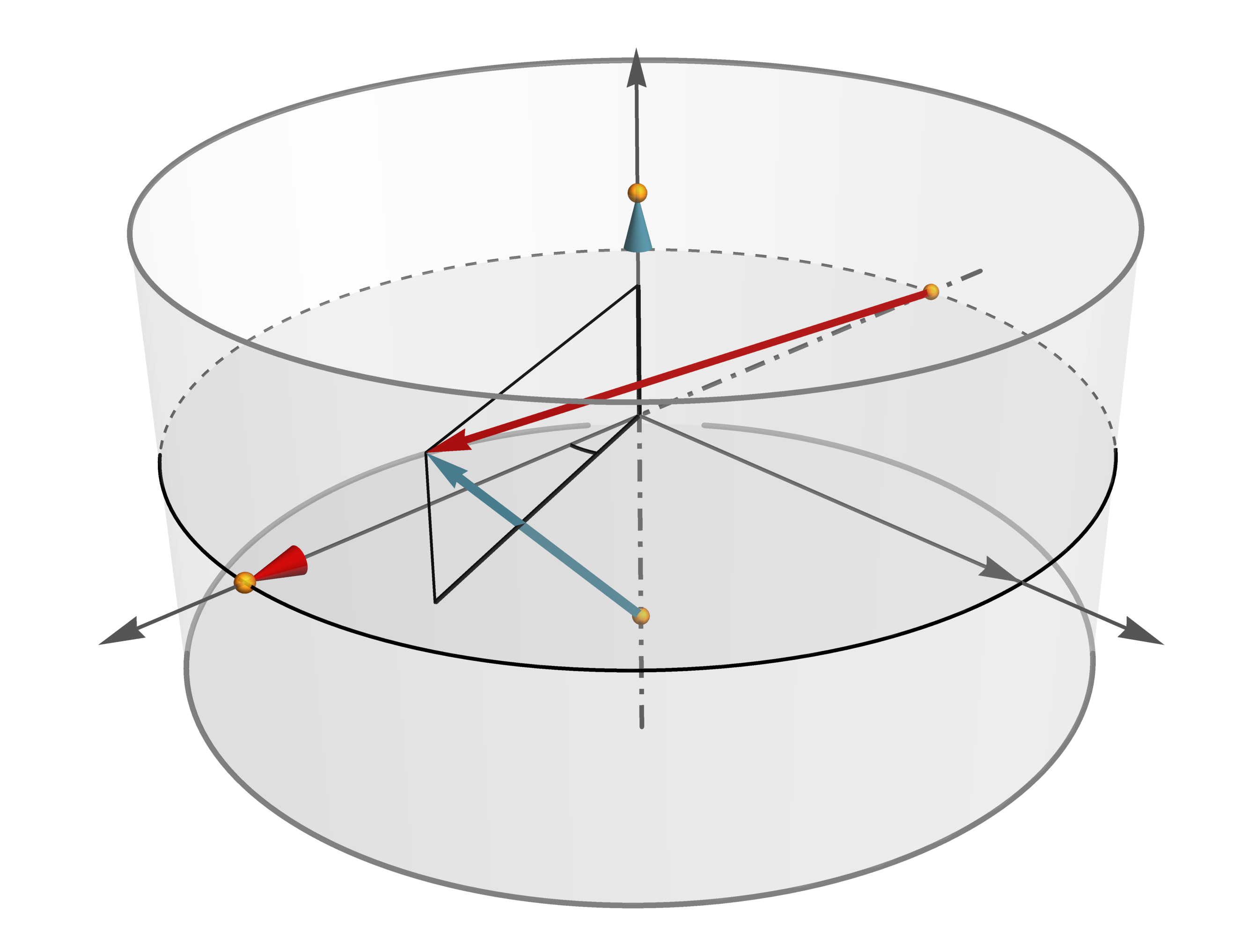}
\put(18,25){$\mathbf{x}_1^{\prime\prime}$}
\put(74,55.5){$\mathbf{x}_1^{\prime}$}
\put(2,24){$x_{\mathrm{p}1}$}
\put(95,24){$x_{\mathrm{p}2}$}
\put(47,75){$x_{\mathrm{p}3}$$=$$z_{\mathrm{p}}$}
\put(52,41){$\mathbf{x}_0^{\prime}$}
\put(53,25){$\mathbf{x}_3^{\prime}$}
\put(46,61){$\mathbf{x}_3^{\prime\prime}$}
\put(42,37){$\phi$}
\put(32,41){$\mathbf{x}$}
\put(58,51){$\mathbf{r}_1$}
\put(42,29){$\mathbf{r}_3$}
\put(32,27){$r_{\mathrm{p}}$}
\put(40,10){cylinder: $V^{\prime}_\mathrm{c}$}
\put(75,29){$r_{\mathrm{c}}$}
\end{overpic}\caption{\label{fig-3} Cylinder geometry with the schematic of a transverse displacement along $x_{\mathrm{p}1}$ ($n$=1) and that along the axial direction of $x_{\mathrm{p}3}$ ($n$=3), respectively, required in the cylindrical regularization.}
\end{figure}

Since the disk shape is a special case of cylindrical geometry, flattening a cylinder can be considered to estimate the characteristic matrix elements $\mathcal{D}_{\mathrm{d}nn}$ of the discal regularization. In a cylinder $V^{\prime}_{\mathrm{c}}$, we have $\mathcal{D}_{\mathrm{c}11}$=$\mathcal{D}_{\mathrm{c}22}$ due to the rotational symmetry. Along the two principal axis $x_{\mathrm{p}1,\mathrm{p}3}$, the diagonal element $\mathcal{D}_{\mathrm{c}11}$ and $\mathcal{D}_{\mathrm{c}33}$ can be calculated according to the geometry in Fig.~\ref{fig-3}. $\mathcal{D}_{\mathrm{c}11}$ along the displacement $\mathbf{r}^{\prime}_1$=$\mathbf{x}_1^{\prime\prime}$$-$$\mathbf{x}_1^{\prime}$ can be expressed in the cylindrical coordinates ($r_{\mathrm{p}}$, $\phi$, $z_{\mathrm{p}}$=$x_{\mathrm{p}3}$) as
\begin{align}
\mathcal{D}_{\mathrm{c}11}=&\frac{1}{2\pi r^{\prime}_1}\int_{V^{\prime}_{\mathrm{c}}}\frac{r_{1\parallel}}{r^3_1}
\mathrm{d}^3x=\frac{1}{4\pi r_{\mathrm{c}}}\int_{V^{\prime}_{\mathrm{c}}}\frac{r_{1\parallel}}{r^3_1}
r_{\mathrm{p}}\mathrm{d}r_{\mathrm{p}}\mathrm{d}\phi\mathrm{d}z_{\mathrm{p}}\nonumber\\
=&\frac{1}{4\pi}\int_{-\kappa}^{\kappa}\mathrm{d}\zeta\int_0^1\mathrm{d}\xi\int_0^{2\pi}\frac{(1+\xi\cos\phi)\xi\mathrm{d}\phi}{(1+\zeta^2+\xi^2+2\xi\cos\phi)^{3/2}}\nonumber\\
&(r^{\prime}_1=2r_{\mathrm{c}},\,\xi=r_{\mathrm{p}}/r_{\mathrm{c}},\,\zeta=z_{\mathrm{p}}/r_{\mathrm{c}})\,,
\label{Appb-C-Regularization-D11}
\end{align}
where the thickness of the cylinder is set as $l$=$\kappa r^{\prime}_1$ by an adjustable parameter $\kappa$. The integral in Eq.~(\ref{Appb-C-Regularization-D11}) is hard to solve in a general case. However, the discal matrix elements result from a limit $\kappa$$\rightarrow$0, i.e.,
\begin{align}
\mathcal{D}_{\mathrm{c}11}<&\frac{\kappa}{2\pi}\int_0^1\mathrm{d}\xi\int_0^{2\pi}\frac{(1+\xi\cos\phi)\xi\mathrm{d}\phi}{(1+\xi^2+2\xi\cos\phi)^{3/2}}\nonumber\\
<&\frac{\kappa}{2\pi}\int_0^1\mathrm{d}\xi\int_0^{2\pi}\frac{\xi\mathrm{d}\phi}{(1+\xi^2+2\xi\cos\phi)^{1/2}}\nonumber\\
=&\frac{2\kappa}{\pi}\rightarrow\mathcal{D}_{\mathrm{d}11}=0\,\mathrm{with}\,\kappa\rightarrow0\,,
\label{D11-Estimation}
\end{align}
where the integral in the second line is equivalent to the electric potential on the rim of a uniformly charged disk of a unit length of radius, with its charge area density $\kappa\epsilon_0$ (see Sec. 2.6 in Ref.~\cite{Purcell}). The calculation of $\mathcal{D}_{\mathrm{c}33}$ along $\mathbf{r}^{\prime}_3$=$\mathbf{x}_3^{\prime\prime}$$-$$\mathbf{x}_3^{\prime}$ is similar to that in Eq.~(\ref{Appa-Dirac-Integration1}); it is given by, with $r^{\prime}_3$=$\kappa r_{\mathrm{c}}$,
\begin{align}
\mathcal{D}_{\mathrm{c}33}=&\frac{1}{2\pi r^{\prime}_3}\int_{V^{\prime}_{\mathrm{c}}}\frac{r_{3\parallel}}{r^3_3}
\mathrm{d}^3x=\frac{1}{2\pi r^{\prime}_3}\int_{V^{\prime}_{\mathrm{c}}}\frac{r_{3\parallel}}{r^3_3}
r_{\mathrm{p}}\mathrm{d}r_{\mathrm{p}}\mathrm{d}\phi\mathrm{d}z_{\mathrm{p}}\nonumber\\
=&\int_{0}^{1}\mathrm{d}\zeta\int_0^{\frac{1}{\kappa}}\frac{\zeta\xi\mathrm{d}\xi}{(\zeta^2+\xi^2)^{3/2}}
\left(\xi=\frac{r_{\mathrm{p}}}{r^{\prime}_3},\,\zeta=\frac{1}{2}+\frac{z_{\mathrm{p}}}{r^{\prime}_3}\right)\nonumber\\
=&\frac{1+\kappa-\sqrt{1+\kappa^2}}{\kappa}\rightarrow\mathcal{D}_{\mathrm{d}33}=1\,\mathrm{with}\,\kappa\rightarrow0\,.
\label{Appb-C-Regularization-D33}
\end{align}
It shows that only the longitudinal component $\mathcal{D}_{\mathrm{d}33}$ is nonzero in the discal regularization. Setting $\mathbf{\hat{s}}$ as a unit normal vector of the disk along the $x_{\mathrm{p}3}$-axis, we have $\bm{\mathcal{D}}_{\mathrm{d}}$=$\mathbf{\hat{s}}\mathbf{\hat{s}}$, which is in agreement with Eq.~(\ref{Local-DC-Limit-2}) and gives rise to Eq.~(\ref{Disc-Regularization}).

Actually, the dipole distribution in other spaces with different dimensions still carries a similar singularity. For example, in the 2-D space, it can be checked that
\begin{align}
\nabla\nabla \ln(r)=-\left(\frac{2\mathbf{\hat{r}}\mathbf{\hat{r}}-\bm{\mathcal{I}}}{r^2}\right)_{\!\!\bar{V}^{\prime}_{\varepsilon}}+2\pi\bm{\mathcal{D}}_{\varepsilon}\delta(\mathbf{r})\,.
\label{Line-Regularization}
\end{align}
In this case, the typical discal regularization is degenerated into a line regularization, where $\bm{\mathcal{D}}_{\varepsilon}$=$\mathbf{\hat{s}}\mathbf{\hat{s}}$, and $\mathbf{\hat{s}}$ is the unit vector normal to an infinitesimal needle-like region.

In general, when an integration over an $n$-dimensional space is scale-invariant, a shape-dependent singularity may occur. Let the following integral
\begin{align}
I_{i_1\cdots i_k}=\int_{V_1}\mathcal{F}_{i_1\cdots i_k}(\mathbf{r})\mathrm{d}^nx\,,i_{j=1,\cdots,k}=1,\cdots,n
\label{Scaling-Invariant-Integration}
\end{align}
be bounded by $0$$<$$|I_{\cdots}|$$<$$\infty$, where $\mathcal{F}_{\cdots}$ is the components of a $k$th-order tensor $\bm{\mathcal{F}}$, and $V_1$ is the integration region including the origin $\bm{o}$=$\mathbf{x}^{\prime}$. If this integral is scale-invariant, i.e.,
\begin{align}
\mathcal{F}_{i_1\cdots i_k}(b\mathbf{r})=b^{-n}\mathcal{F}_{i_1\cdots i_k}(\mathbf{r})\,,
\label{Scaling-Rule}
\end{align}
for an arbitrary positive number $b$. Then, the integration region becomes $V(b)$, which is reduced by $b$ times compared with $V_1$=$V(b=1)$. Obviously, $\bm{\mathcal{F}}$ is not defined at $\bm{o}$, so it requires regularization. In a limit process $b$$\rightarrow$$\infty$, $V(b\rightarrow\infty)$ can be taken as an infinitesimal neighborhood around $\bm{o}$: $\delta V$. However, $I_{\cdots}$ remains unchanged in this limit process $V_1$$\rightarrow$$\delta V$ because the integral function does not depend on $b$, as indicated by $\mathrm{d}^n(bx)$=$b^n\mathrm{d}^nx$. It suggests that
\begin{align}
\mathcal{F}_{i_1\cdots i_k}(\mathbf{r})|_{\delta V}\simeq I_{i_1\cdots i_k}\delta(\mathbf{r})\,,
\label{Equivalence-delta1}
\end{align}
where $I_{\cdots}$ is related to the shape of $V_1$ or $\delta V$. It confirms that there is definitely a singularity at $\bm{o}$. In general, an arbitrary regularization expresses the tensor function within a infinitesimal geometry $V^{\prime}_{\varepsilon}$ as
\begin{align}
\mathcal{F}_{i_1\cdots i_k}(\mathbf{r})|_{V^{\prime}_{\varepsilon}}\simeq a_{i_1\cdots i_k}(\varepsilon)\delta(\mathbf{r})\,.
\label{Equivalence-delta2}
\end{align}
When $\bm{\mathcal{F}}$ is defined by the derivative of a ($k$$-$1)th tensor $\bm{\mathcal{G}}$ as
\begin{align}
\mathcal{F}_{i_1\cdots i_k}(\mathbf{r})|_{\bar{V}^{\prime}_{\varepsilon}}=\partial^{\prime}_{i_k}\mathcal{G}_{i_1\cdots i_{k-1}}(\mathbf{r})|_{\bar{V}^{\prime}_{\varepsilon}}\,,
\label{G-Tensor}
\end{align}
the difference-quotient method can be used to determine the characteristic tensor $a_{\cdots}(\varepsilon)$ of its internal distribution, where the derivative $\partial^{\prime}_{i_k}$ acts with respective to the variable $x^{\prime}_{i_k}$. Evidently, $\bm{\mathcal{G}}$ itself is a singular distribution.

\section{Some integral calculations associated with the dipole distribution}
First, the calculations of the integrals in Eqs.~(\ref{Vector-Potential-Coulomb-2}) and (\ref{QSteady-Magnetic-Field}) are detailed. Next, some hints and intermediate steps in the calculation of the magnetic field from the instantaneous Biot-Savart integral are presented.

The last integral in Eq.~(\ref{Vector-Potential-Coulomb-2}) is
\begin{align}
\int_{\bar{V}^{\prime\prime}_{\varepsilon}}\frac{3\mathbf{\hat{r}}^{\prime}\mathbf{\hat{r}}^{\prime}-\bm{\mathcal{I}}}{4\pi rr^{\prime3}}\cdot\mathbf{J}\left(\mathbf{x}^{\prime\prime},t_r\right)\mathrm{d}^3x^{\prime}\,.
\label{Appc-Part-Dipole-Integration-1}
\end{align}
\begin{figure}[ht]
\centering
\begin{overpic}
[scale=1.25]{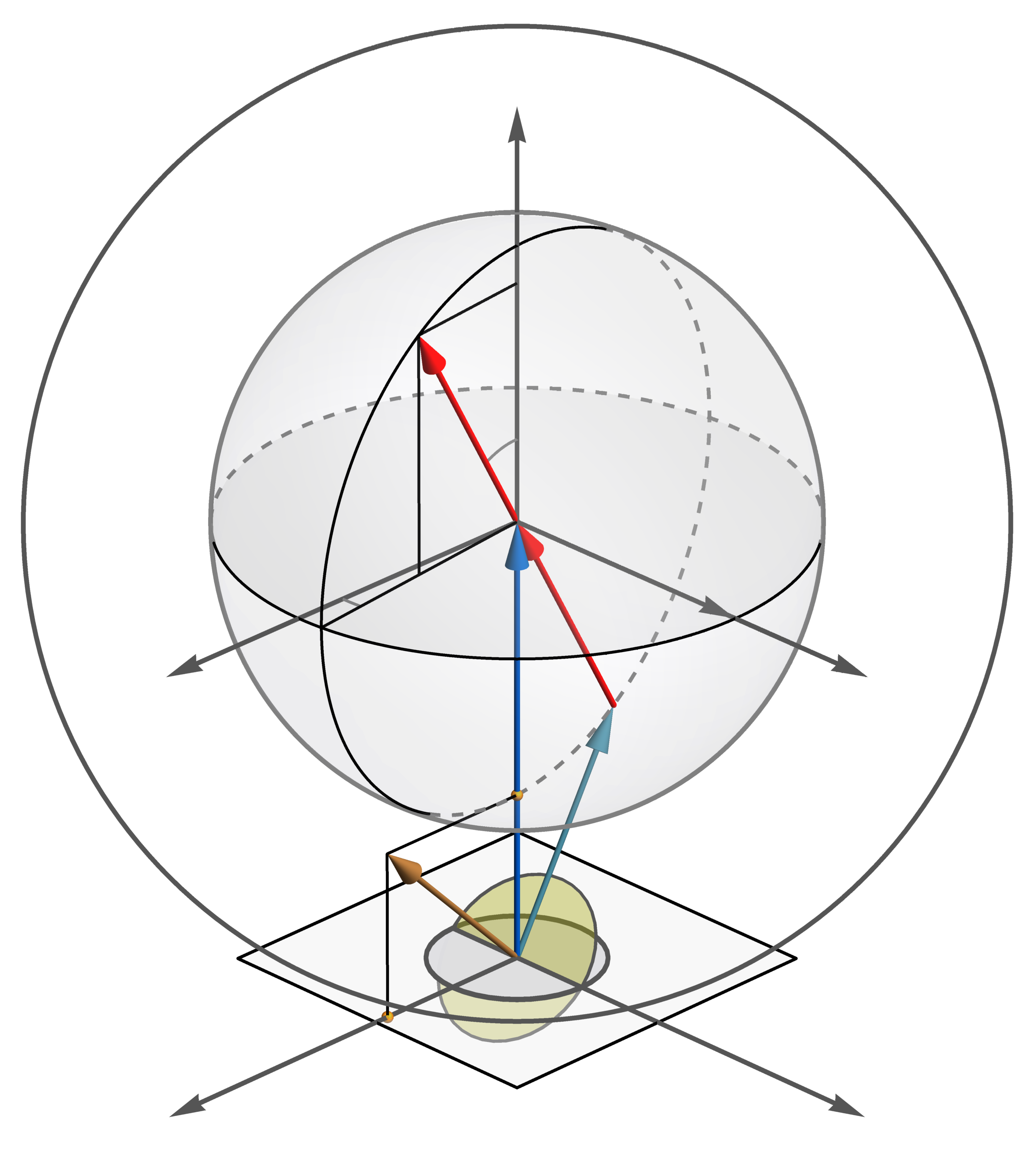}
\put(11,56){$S(r)$}
\put(11,40){$x_1$}
\put(75,40){$x_2$}
\put(42.5,92){$x_3$}
\put(13,6){$\mathbf{\hat{e}}_1$}
\put(72.5,6){$\mathbf{\hat{e}}_2$}
\put(33.5,72){$\mathbf{\bar{x}}^{\prime}$}
\put(53,38){$\mathbf{x}^{\prime}$}
\put(43,14.5){$\mathbf{x}^{\prime\prime}$}
\put(69,80){$S(R)$}
\put(51,21){$V^{\prime\prime}_\mathrm{d}$}
\put(45,56){$\mathbf{x}$}
\put(38,61){$\mathbf{r}$}
\put(50,46){$\mathbf{r}$}
\put(51,31){$\mathbf{r}^{\prime}$}
\put(40,40){$\mathbf{R}$}
\put(30,46){$\phi$}
\put(41.5,62){$\theta$}
\put(56,51){$r$}
\put(31,8){$\mathbf{J}^{\prime\prime}_1$}
\put(40,33){$\mathbf{J}^{\prime\prime}_3$}
\put(29,28){$\mathbf{J}^{\prime\prime}_{\mathrm{ret}}$}
\put(33.3,18.0){$V_{\varepsilon}^{\prime\prime}$}
\put(34,3){Tangent Plane}
\end{overpic}\caption{\label{fig-4} Two concentric sphere surfaces $S(r)$ and $S(R)$ shown in the spherical coordinate frame, centered at $\mathbf{x}$, with a tilted disk $V_{\varepsilon}^{\prime\prime}$ and a normal disk $V_{\mathrm{d}}^{\prime\prime}$ lying in the tangent plane to the sphere $S(R)$ at $\mathbf{x}^{\prime\prime}$, where the current is oriented normal to the $x_2$ axis.}
\end{figure}
The configuration of the spherical coordinate system with the coordinate origin $\mathbf{x}$ and the source point $\mathbf{x}^{\prime\prime}$ is shown in Fig. \ref{fig-4}, where the current can be formally simplified as $\mathbf{J}^{\prime\prime}_{\mathrm{ret}}$=$\mathbf{J}(\mathbf{x}^{\prime\prime},t_r)$, oriented on the $x_1$$-$$x_3$ plane. The angular integrations with respect to ($\theta$, $\phi$) are primarily performed on a spherical surface $S(r)$ with the radius $r$. For a better illustration, the endpoint of vector $\mathbf{r}$ is shifted from the point $\mathbf{x}$ to the antipodal point $\mathbf{\bar{x}}^{\prime}$ of $\mathbf{x}^{\prime}$ in parallel, which does not vary the integration on $S(r)$. A default normal discal regularization is not suitable here because the given disk $V_{\mathrm{d}}^{\prime\prime}$ complicates the angular integrations. Instead, an effective alternative is a specified tilted-disk regularization, where the disk $V_{\varepsilon}^{\prime\prime}$ lies on the tangent plane to the spherical surface $S(R)$ at the point $\mathbf{x}^{\prime\prime}$. Moreover, the integration region $\bar{V}_{\varepsilon^{\prime\prime}}$ can be equivalently replaced by $V_{\infty}$ without $S(R)$.

According to the geometry in Fig.~\ref{fig-4}, the relation $\mathbf{r}^{\prime}$+$\mathbf{r}$=$\mathbf{R}$ implies
\begin{subequations}\label{Appc-Some-Relations-1}
\begin{align}
r^{\prime2}=&R^2+r^2-2rR\cos\theta\,,\\
\mathbf{\hat{r}}^{\prime}=&\,r^{\prime-1}(\mathbf{R}-\mathbf{r})\,,\\
\mathbf{\hat{r}}^{\prime}\mathbf{\hat{r}}^{\prime}=&\,r^{\prime-2}(\mathbf{R}-\mathbf{r})(\mathbf{R}-\mathbf{r})\,.
\end{align}
\end{subequations}
In Eqs.~(\ref{Appc-Some-Relations-1}), owing to the axial ($x_3$) symmetry of $\mathbf{r}^{\prime}$, the integrals involving $\sin\phi$, $\cos\phi$ or $\sin\phi\cos\phi$ vanish and only those proportional to the dyad components $\mathbf{\hat{e}}_{i}\mathbf{\hat{e}}_{i}$ ($i$=1, 2, 3) are nonzero. Therefore, the following dyad components in $\mathbf{\hat{r}}^{\prime}\mathbf{\hat{r}}^{\prime}$
\begin{align}
&\sum_{i=1}^3(\mathbf{\hat{r}}^{\prime}\mathbf{\hat{r}}^{\prime}:\mathbf{\hat{e}}_{i}\mathbf{\hat{e}}_{i})\mathbf{\hat{e}}_{i}\mathbf{\hat{e}}_{i}\nonumber\\
=&\frac{(r\sin\theta\cos\phi)^2\mathbf{\hat{e}}_1\mathbf{\hat{e}}_1+(R-r\cos\theta)^2\mathbf{\hat{e}}_3\mathbf{\hat{e}}_3}{R^2+r^2-2rR\cos\theta}\nonumber\\
+&\frac{(r\sin\theta\sin\phi)^2\mathbf{\hat{e}}_2\mathbf{\hat{e}}_2}{R^2+r^2-2rR\cos\theta}
\label{Appc-Effective-Dyad}
\end{align}
should be considered in integral~(\ref{Appc-Part-Dipole-Integration-1}). Since the current is normal to $\mathbf{\hat{e}}_2$, it can be projected onto the other two directions as $\mathbf{J}^{\prime\prime}_3$=$\mathbf{\hat{e}}_3\mathbf{\hat{e}}_3\cdot\mathbf{J}^{\prime\prime}_{\mathrm{ret}}$ and $\mathbf{J}^{\prime\prime}_1$=$\mathbf{\hat{e}}_1\mathbf{\hat{e}}_1\cdot\mathbf{J}^{\prime\prime}_{\mathrm{ret}}$, respectively. This also indicates that only the second line in Eq.~(\ref{Appc-Effective-Dyad}) remains nonzero. Hence, the differential form of (\ref{Appc-Part-Dipole-Integration-1}) can be simplified as
\begin{align}
\left[\mathbf{J}^{\prime\prime}_3 I_1(\mathit{\Theta})+\mathbf{J}^{\prime\prime}_1 I_2(\mathit{\Theta},\phi)\right]\frac{\eta\mathrm{d}V}{4\pi R}
\label{Appc-Part-Dipole-Integration-e1-e3}
\end{align}
with the angular terms
\begin{align}
I_1=\frac{3(1+\eta\mathit{\Theta})^2}{(1+\eta^2+2\eta\mathit{\Theta})^{5/2}}-\frac{1}{(1+\eta^2+2\eta\mathit{\Theta})^{3/2}}
\label{Appc-I-1}
\end{align}
and
\begin{align}
I_2=\frac{3\eta^2\cos^2\!\phi(1-\mathit{\Theta}^2)}{(1+\eta^2+2\eta\mathit{\Theta})^{5/2}}-\frac{1}{(1+\eta^2+2\eta\mathit{\Theta})^{3/2}}\,,
\label{Appc-I-2}
\end{align}
where $\eta$=$r/R$, $\mathrm{d}V$=$\mathrm{d}\eta\mathrm{d}\mathit{\Theta}\mathrm{d}\phi$, and $\mathit{\Theta}$=$-\cos\theta$ as before. An equivalent region of $\bar{V}^{\prime\prime}_{\mathrm{d}}(\mathbf{\hat{e}}_3)$ is the space restricted by $\eta$$\neq$1, which is all space excluding the surface $S(R)$. It can be verified that
\begin{align}
\int_{0}^{2\pi}\mathrm{d}\phi[I_1(\mathit{\Theta})+2I_2(\mathit{\Theta},\phi)]=0\,,
\label{Appc-I-1-2}
\end{align}
which means that only $I_1$ or $I_2$ needs to be determined. Utilizing the Legendre expansion \cite{Jackson1}
\begin{align}
\frac{R}{r^{\prime}}=\frac{1}{\sqrt{1+\eta^2+2\eta\mathit{\Theta}}}=\sum_{l=0}^{\infty}\frac{\eta_{<}^l}{\eta_{>}^{l+1}}P_l(-\mathit{\Theta})\,,
\label{Appc-Legendre-Expansion}
\end{align}
where $P_l(-\mathit{\Theta})$ is the $l$th order Legendre polynomial, $I_1$ can be expanded as
\begin{align}
I_1=&R^3\frac{\partial^2}{\partial R^2}\left(\frac{1}{r^{\prime}}\right)\nonumber\\
=&\sum_{l=0}^{\infty}\frac{\eta_{<}^l}{\eta_{>}^{l+3}}(l+1)(l+2)
\begin{cases}
P_l(-\mathit{\Theta})\,,\quad\eta<1\\
P_{l+2}(-\mathit{\Theta})\,,\eta>1
\end{cases}\!\!.
\label{Appc-Legendre-Expansion-I-1}
\end{align}
The angular integrations of Eq.~(\ref{Appc-Legendre-Expansion-I-1}) only respect the zeroth order term and becomes
\begin{align}
\int_0^{2\pi}\mathrm{d}\phi\int_{-1}^1\mathrm{d}\mathit{\Theta} I_1(\mathit{\Theta})=4\pi[1+\mathrm{sign}(1-\eta)]\,,
\label{Appc-Integration-I-1}
\end{align}
where $\mathrm{sign}(\xi)$ is a typical piecewise function, as is $-$1, 0, or 1 depending on whether $\xi$ is negative, zero, or positive.
Equations~(\ref{Appc-Part-Dipole-Integration-e1-e3}), (\ref{Appc-I-1-2}) and(\ref{Appc-Integration-I-1}) resolve Eq.~(\ref{Appc-Part-Dipole-Integration-1}) to
\begin{align}
\int_0^1&\!\mathrm{d}\eta\frac{2\mathbf{J}^{\prime\prime}_3-\mathbf{J}^{\prime\prime}_1}{2R}\eta\left[1+\mathrm{sign}(1-\eta)\right]\nonumber\\
=&\frac{3\mathbf{\hat{R}}\mathbf{\hat{R}}-\bm{\mathcal{I}}}{R^3}\cdot\!\int_0^{R}\!\mathbf{J}(\mathbf{x}^{\prime\prime},t_r)
r\mathrm{d}r\nonumber\\
=&c^2\frac{3\mathbf{\hat{R}}\mathbf{\hat{R}}-\bm{\mathcal{I}}}{R^3}\cdot\int_{t-R/c}^{t}\mathbf{J}(\mathbf{x}^{\prime\prime},\tau)(t-\tau)\mathrm{d}\tau\,.
\label{Appc-Part-Dipole-Integration-2}
\end{align}
The first line in Eq.~(\ref{Appc-Part-Dipole-Integration-2}) indicates that the integration in the region $\eta$$>$1 is zero, which contracts the effective integral to the region 0$\leqslant$$\eta$$<$1. In terms of the first paragraph in Appendix B, the unit vector normal to the surface can be taken as $\mathbf{\hat{s}}$=$\pm\mathbf{\hat{R}}$ that prompts $\bm{\mathcal{E}}$=$\bm{\mathcal{I}}$$-$$\mathbf{\hat{R}}\mathbf{\hat{R}}$. Equation~(\ref{Appc-Part-Dipole-Integration-2}) directly leads to the vector potential given in Eq.~({\ref{Vector-Potential-Coulomb-3}}) where the point $\mathbf{x}^{\prime\prime}$ is replaced by $\mathbf{x}^{\prime}$ for consistency in the formulas, which changes $\mathbf{R}$ into $\mathbf{r}$.

Now, let's turn to the last integral in Eq.~(\ref{QSteady-Magnetic-Field}), i.e.,
\begin{align}
\int_{\bar{V}^{\prime\prime}_{\varepsilon}}\frac{(3\mathbf{\hat{r}}^{\prime}\mathbf{\hat{r}}^{\prime}-\bm{\mathcal{I}})\cdot\mathbf{J}(\mathbf{x}^{\prime\prime})}{4\pi r^2r^{\prime3}}\times\mathbf{\hat{r}}\mathrm{d}^3x^{\prime}\,.
\label{Appc-Part-Dipole-Integration-3}
\end{align}
Figure \ref{fig-4} is still suitable for this case, where the current can be formally simplified as $\mathbf{J}^{\prime\prime}$=$\mathbf{J}(\mathbf{x}^{\prime\prime})$, oriented in the $x_1$$-$$x_3$ plane as well. Similarly, $\mathbf{J}^{\prime\prime}$ does not have a $\mathbf{\hat{e}}_2$ component, so the factor $(\mathbf{\hat{r}}^{\prime}\mathbf{\hat{r}}^{\prime}\cdot\mathbf{J}^{\prime\prime})$$\times\mathbf{\hat{r}}$ can be expressed by Eqs.~(\ref{Appc-Some-Relations-1}) as
\begin{align}\label{Appc-Part-Dipole-Integration-3-1}
&(\mathbf{\hat{r}}^{\prime}\cdot\mathbf{J}^{\prime\prime})\mathbf{\hat{r}}^{\prime}\times\mathbf{\hat{r}}
=\frac{R}{r^{\prime2}}[\mathbf{J}^{\prime\prime}\cdot(\mathbf{R}-\mathbf{r})]\mathbf{\hat{e}}_3\times\mathbf{\hat{r}}\nonumber\\
=&\frac{R}{r^{\prime2}}[\mathbf{J}^{\prime\prime}\cdot(\mathbf{R}-\mathbf{r})]
(\mathbf{\hat{r}}\cdot\mathbf{\hat{e}}_1\mathbf{\hat{e}}_2-\mathbf{\hat{r}}\cdot\mathbf{\hat{e}}_2\mathbf{\hat{e}}_1)\nonumber\\
=&-\frac{r R}{r^{\prime2}}(\mathbf{J}^{\prime\prime}\cdot\mathbf{\hat{e}}_1)(\mathbf{\hat{r}}\cdot\mathbf{\hat{e}}_1)^2\mathbf{\hat{e}}_2+\mathrm{other\,terms}\,.
\end{align}
In Eq.~(\ref{Appc-Part-Dipole-Integration-3-1}), only the term proportional to $(\mathbf{\hat{r}}\cdot\mathbf{\hat{e}}_1)^2$ can survive due to the axial symmetry of $r^{\prime}$. The other term $(\bm{\mathcal{I}}\cdot\mathbf{J}^{\prime\prime})\times\mathbf{\hat{r}}$, i.e., $\mathbf{J}^{\prime\prime}\times\mathbf{\hat{r}}$ can be decomposed into
\begin{align}\label{Appc-Part-Dipole-Integration-3-2}
\mathbf{J}^{\prime\prime}\times\mathbf{\hat{r}}=&(\mathbf{J}^{\prime\prime}\cdot\mathbf{\hat{e}}_1)
[(\mathbf{\hat{r}}\cdot\mathbf{\hat{e}}_2)\mathbf{\hat{e}}_3-(\mathbf{\hat{r}}\cdot\mathbf{\hat{e}}_3)\mathbf{\hat{e}}_2]\nonumber\\
+&(\mathbf{J}^{\prime\prime}\cdot\mathbf{\hat{e}}_3)
[(\mathbf{\hat{r}}\cdot\mathbf{\hat{e}}_1)\mathbf{\hat{e}}_2-(\mathbf{\hat{r}}\cdot\mathbf{\hat{e}}_2)\mathbf{\hat{e}}_1]\,,
\end{align}
all of which vanish in the integration except for the term including $\mathbf{\hat{r}}\cdot\mathbf{\hat{e}}_3$. Thus, the effective differential in (\ref{Appc-Part-Dipole-Integration-3}) can be rewritten as
\begin{align}\label{Appc-Part-Dipole-Integration-3-3}
-\frac{I_3(\mathit{\Theta},\phi)\mathrm{d}V}{4\pi R^2}\mathbf{J}^{\prime\prime}\cdot\mathbf{\hat{e}}_1\mathbf{\hat{e}}_2=\frac{I_3(\mathit{\Theta},\phi)\mathrm{d}V}{4\pi R^2}\mathbf{J}^{\prime\prime}\times\mathbf{\hat{R}}
\end{align}
with
\begin{align}\label{Appc-I-3}
I_3=\frac{3\eta\cos^2\!\phi(1-\mathit{\Theta}^2)}{(1+\eta^2+2\eta\mathit{\Theta})^{5/2}}
+\frac{\mathit{\Theta}}{(1+\eta^2+2\eta\mathit{\Theta})^{3/2}}\,.
\end{align}
By comparing Eq.~(\ref{Appc-I-3}) with Eq.~(\ref{Appc-I-2}), a relation
\begin{align}\label{Appc-I-R}
\eta I_3=I_2-R^2\frac{\partial}{\partial R}\left(\frac{1}{r^{\prime}}\right)
\end{align}
can be identified, where the two terms on the rhs., nevertheless, cancel out in the angular integrations using the Legendre expansion (\ref{Appc-Legendre-Expansion}). This also causes the integral (\ref{Appc-Part-Dipole-Integration-3-3}) in the second line of Eq.~(\ref{QSteady-Magnetic-Field}) to vanish. In this case, the expression $\bm{\mathcal{E}}$=$\bm{\mathcal{I}}$$-$$\mathbf{\hat{R}}\mathbf{\hat{R}}$ still holds.

Finally, a direct calculation of the magnetic field from the instantaneous Biot-Savart integral will be considered below. By use of Eq.~(\ref{R-DC-General-Distribution-2}) with $t^{\prime}_r$=$t$$-$$r^{\prime}/c$, Eq.~(\ref{New-H}) can be expanded as
\begin{align}
\label{Appc-New-H-Expansion}
\mathbf{H}=&\int_{V_{\infty}}\frac{[\bm{\mathcal{E}}\cdot\mathbf{J}(\mathbf{x}^{\prime},t)]\times\mathbf{\hat{r}}}{4\pi r^2}
\mathrm{d}^3x^{\prime}\nonumber\\
+&\int_{V_{\infty}}\!\!\frac{\mathrm{d}^3x^{\prime\prime}}{4\pi}\left(\int_{\bar{V}^{\prime\prime}_{\varepsilon}}\!\!\mathrm{d}^3x^{\prime}\left\{\left[\frac{\mathbf{J}(\mathbf{x}^{\prime\prime},t^{\prime}_r)}
{4\pi r^2r^{\prime3}}+\frac{\dot{\mathbf{J}}(\mathbf{x}^{\prime\prime},t^{\prime}_r)}{4\pi cr^2r^{\prime2}}\right]\right.\right.\nonumber\\
\cdot&(3\mathbf{\hat{r}^{\prime}}\mathbf{\hat{r}^{\prime}}-\bm{\mathcal{I}})
+\left.\left.\frac{\mathbf{\hat{r}^{\prime}}\times\mathbf{\hat{r}^{\prime}}\times\ddot{\mathbf{J}}(\mathbf{x}^{\prime\prime},t^{\prime}_r)}{4\pi c^2r^2r^{\prime}}\right\}\times\mathbf{\hat{r}}\right)\,,
\end{align}
where the order of integration is interchanged. The integration with respective to $\mathbf{x}^{\prime}$ in this expression can be calculated through a series of cumbersome steps in the spherical coordinate system, which is omitted here except that some hints are provided here. Equation~(\ref{Appc-New-H-Expansion}) demonstrates that the magnetic field at $\mathbf{x}$ is induced instantaneously by the displacement current at $\mathbf{x}^{\prime}$, which, nevertheless, is related to the current at $\mathbf{x}^{\prime\prime}$ with a retardation. Perversely in Eq.~(\ref{Appc-New-H-Expansion}), the final contributions of the displacement current over all $\mathbf{x}^{\prime}$ come from $\dot{\mathbf{J}}^{\prime\prime}_{\mathrm{ret}}$ and $\ddot{\mathbf{J}}^{\prime\prime}_{\mathrm{ret}}$, while those from $\mathbf{J}^{\prime}$ and $\mathbf{J}^{\prime\prime}_{\mathrm{ret}}$ are canceled. A final simplification of Eq.~(\ref{Appc-New-H-Expansion}) becomes Jefimenko's equation (\ref{J-H-Equation}) as expected.

For the integration in Eq.~(\ref{Appc-New-H-Expansion}), it should be considered in another spherical coordinate system which takes $\mathbf{x}^{\prime\prime}$ as the coordinate origin with a vector integral variable of $\mathbf{r}^{\prime}$. Thus, $S(r)$ should be replaced by $S(r^{\prime})$, and $\mathbf{x}^{\prime\prime}$ is the mutual center of $S(r^{\prime})$ and $S(R)$. With this alteration in terms of Fig.~\ref{fig-4}, an exchange between $\mathbf{r}$ and $\mathbf{r}^{\prime}$ should be taken. In this case, the spherical regularization is a suitable one to calculate the integral, which naturally requires $\bm{\mathcal{E}}$=$2\bm{\mathcal{I}}/3$. A few of related expressions that can be expanded using Legendre polynomials are

\begin{align}\label{Appc-I-045}
I_0=&-R^2\frac{\partial}{\partial R}\left(\frac{1}{r}\right)\,,I_4=-\frac{\partial}{\partial\eta}\left(\frac{R}{r}\right)\,,\nonumber\\
I_5=&\frac{1+2\eta\mathit{\Theta}+\eta^2}{3\eta^2}\left(I_2+\frac{I_0-\eta I_4}{1-\eta^2}\right)\,,
\end{align}
where $\eta$=$r^{\prime}/R$, and $I_2$ has the same form as in Eq.~(\ref{Appc-I-2}). Thus, the double integral terms in Eq.~(\ref{Appc-New-H-Expansion}) can be simplified as
\begin{subequations}\label{Appc-J-3}
\begin{align}
(3I_5-I_0)&\mathbf{J}(\mathbf{x}^{\prime\prime},t-\eta R/c)\times\mathbf{\hat{R}}\frac{\mathrm{d}V}{4\pi\eta R^2}\,,\\
\frac{R\eta}{c}(3I_5-I_0)&\dot{\mathbf{J}}(\mathbf{x}^{\prime\prime},t-\eta R/c)\times\mathbf{\hat{R}}\frac{\mathrm{d}V}{4\pi\eta R^2}\,,\\
\left(\frac{R\eta}{c}\right)^{\!2}(I_5-I_0)&\ddot{\mathbf{J}}(\mathbf{x}^{\prime\prime},t-\eta R/c)\times\mathbf{\hat{R}}\frac{\mathrm{d}V}{4\pi\eta R^2}\,,
\end{align}
\end{subequations}
where the three differentials can be integrated with respect to $\mathit{\Theta}$ after the Legendre expansion and with respect to $\eta$ using integration by parts, respectively.
Readers can easily find the intermediate steps in the derivation.
\vspace{1em}
\section{Localization of the external displacement current}
Due to $\nabla\cdot\mathbf{J}$=0 in a quasi-static state, Eq.~(\ref{Longitudinal-Current}) becomes
\begin{align}
\mathbf{J}_{\parallel}
=&\int_{V_{\infty}}\!\bm{\mathcal{L}}_{\varepsilon}(\mathbf{x},\mathbf{x}^{\prime})\cdot\mathbf{J}(\mathbf{x}^{\prime})\mathrm{d}^3x^{\prime}\nonumber\\
=&-\int_{V_{\infty}}\!\mathbf{J}(\mathbf{x}^{\prime})\cdot\nabla^{\prime}\nabla^{\prime}\frac{1}{4\pi r}\mathrm{d}^3x^{\prime}\nonumber\\
=&\int_{V_{\infty}}\!\left(\nabla^{\prime}\frac{1}{4\pi r}\right)\nabla^{\prime}\cdot\mathbf{J}(\mathbf{x}^{\prime})\mathrm{d}^3x^{\prime}\nonumber\\
-&\int_{V_{\infty}}\!\nabla^{\prime}\cdot\left[\mathbf{J}(\mathbf{x}^{\prime})\nabla^{\prime}\frac{1}{4\pi r}\right]\mathrm{d}^3x^{\prime}\nonumber\\
=&-\oint_{S_{\infty}}\!\mathrm{d}\mathbf{s}^{\prime}\cdot\mathbf{J}(\mathbf{x}^{\prime})\nabla^{\prime}\frac{1}{4\pi r}=0\,,
\label{Steady-Parallel-D-Current}
\end{align}
where $S_{\infty}$ is the closed surface of $V_{\infty}$, and the divergence theorem is used in the last line. Since the current is confined to a limited space, the output current through the boundary surface $S_{\infty}$ is zero.

\begin{figure}[ht]
\centering
\begin{overpic}
[scale=0.65]{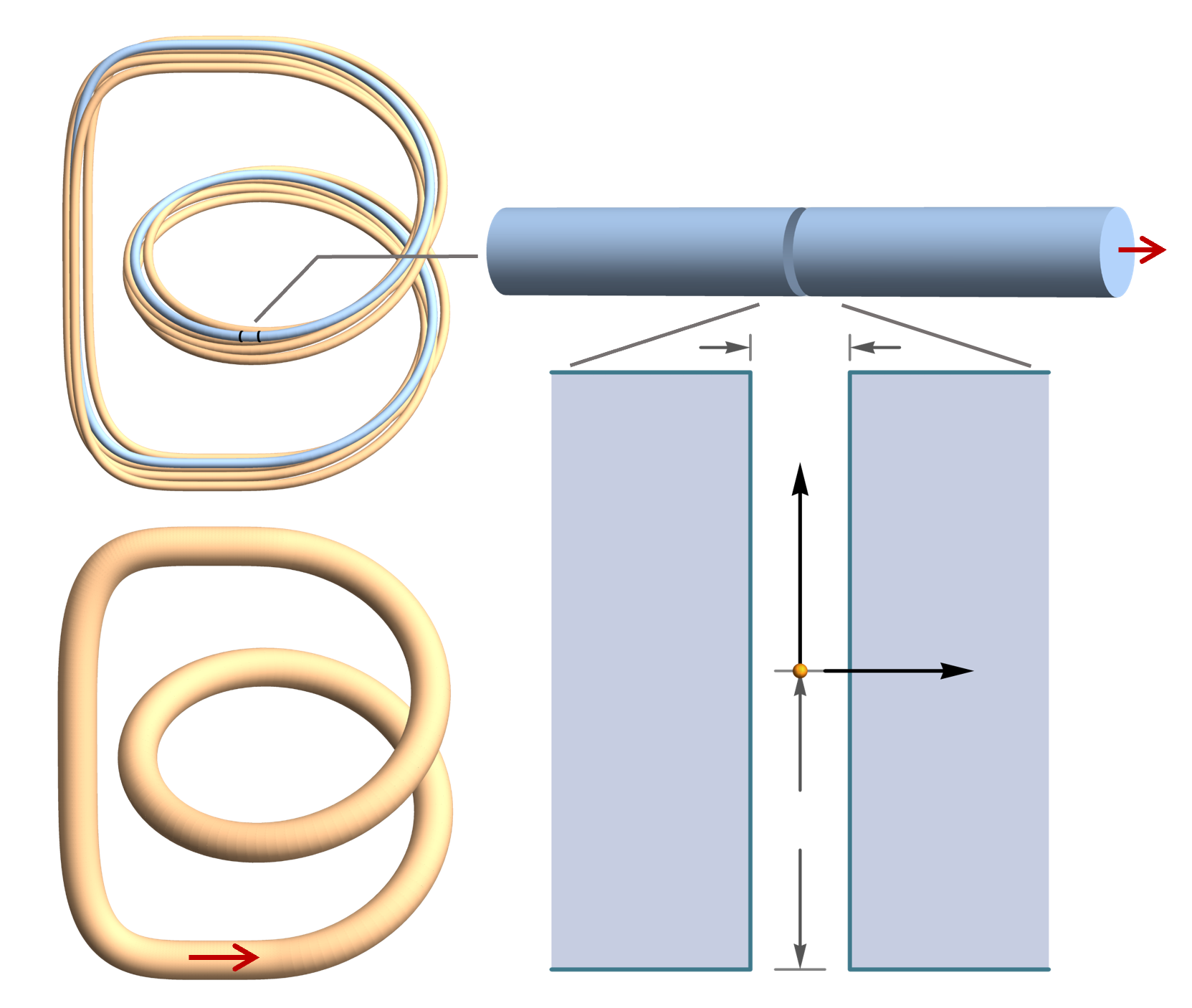}
\put(-2,77){(a)}
\put(66,47){$x^{\prime}_2$}
\put(82.5,27.5){$x^{\prime}_3$}
\put(68,29){$\mathbf{x}$}
\put(64,25){0}
\put(9,72){Closed loops}
\put(65,15){$r_{\mathrm{d}}$}
\put(40,77){(c)}
\put(20,20){$\mathrm{B}$}
\put(14,62){$L_C,C_0$}
\put(20,48){$C_1$}
\put(65,75){$V_\mathrm{d}$}
\put(67,73){\color{gray}\vector(0,-1){6}}
\put(74,70.5){\color{gray}\vector(1,0){20}}
\put(74,70.5){\color{gray}\vector(-1,0){7}}
\put(94,68){\color{gray}\line(0,1){5}}
\put(79,71){$l$}
\put(64.5,54){$2\lambda$}
\put(18,7){$\mathbf{J}^{\prime}$}
\put(-2,35.0){(b)}
\put(48,70){$C_0/V_{\mathrm{d}}$}
\end{overpic}\caption{\label{fig-5} Geometry of (a) closed current loops from (b) a closed circuit `B', with a specified loop $L_C$ distinguished by a different color; and (c) one cylindrical segment $C_0$ within $L_C$ centered at a field point $\mathbf{x}$ with the disk $V_{\mathrm{d}}$ removed, and its enlarged sectional view in the ($x^{\prime}_1$, $x^{\prime}_3$) plane.}
\end{figure}

The result in Eq.~(\ref{Steady-Parallel-D-Current}) can be derived in a straightforward way, which demonstrates how the external displacement current localizes in the current region. A straightforward calculation of Eq.~(\ref{General-Continous-Nonlocal-DC-2}) should naturally be performed using the normal discal regularization ($\bm{\mathcal{E}}$$\simeq$0); however, it requires several steps to refine the integration region. Now, we consider a closed, steady circuit `B', which is regarded as a bundle of closed, cyclic current loops, as shown in Figs.~\ref{fig-5}(a) and \ref{fig-5}(b). Every loop can be structured as a series of interconnected short segments, with their surfaces oriented tangentially to the current direction. A specified current loop $L_C$ is designed to consist of cylindrical segments, one of which, $C_0$, has its center located at the given position $\mathbf{x}$ and a length of $2l$, as illustrated in Fig.~\ref{fig-5}(c). The rest of $L_C$, a long fiber, is labelled with $C_1$. For convenience, a symmetrical disk $V_{\mathrm{d}}$ of radius $r_{\mathrm{d}}$ and thickness $2\lambda$ is chosen in $C_0$, centered at $\mathbf{x}$. The sectional figure of $C_0$ within $\bar{V}_{\mathrm{d}}$ is shown at the bottom of Fig.~\ref{fig-5}(c), where the coordinate system $(x^{\prime}_1, x^{\prime}_2, x^{\prime}_3)$ is established at the point $\mathbf{x}$ to facilitate subsequent calculations.

It will be verified that the contribution to the displacement current at $\mathbf{x}$ from other loops is trivial. Let $X$ be an arbitrary loop that is distinct from $L_C$ in circuit `B'. $X$ is so thin such that a volume integral reduces to a line integral, i.e., $\mathbf{J}(\mathbf{x}^{\prime})\mathrm{d}^3x^{\prime}$=$I_X\mathrm{d}\mathbf{l}$ with the constant current $I_X$, which gives rise to
\begin{align}
\mathbf{J}^{\mathrm{ext}}_{\mathrm{D},X}
=&\int_{X}\!\bm{\mathcal{T}}_{\varepsilon}(\mathbf{x},\mathbf{x}^{\prime})\cdot\mathbf{J}(\mathbf{x}^{\prime})\mathrm{d}^3x^{\prime}
=I_X\oint_{X}\!\mathrm{d}\mathbf{l}\cdot\nabla\nabla\frac{1}{4\pi r}\nonumber\\
=&I_X\sum_{i=1}^{3}\int_{S_X}\!\!(\mathrm{d}\mathbf{s}^{\prime}\times\nabla)\cdot\left(\nabla\nabla\frac{1}{4\pi r}\cdot\mathbf{\hat{e}}_i\right)\mathbf{\hat{e}}_i\nonumber\\
=&I_X\sum_{i=1}^{3}\int_{S_X}\!\!\nabla\times\left(\nabla\nabla\frac{1}{4\pi r}\cdot\mathbf{\hat{e}}_i\right)\cdot\mathrm{d}\mathbf{s}^{\prime}\mathbf{\hat{e}}_i=0,
\label{Appd-Zero-Contribution}
\end{align}
where ${\nabla}^2(1/r)$=0($r$$\neq$0) is used in the second equality, and Stokes' theorem is applied in the third equality with an arbitrary surface $S_X$ covering $X$.
\begin{figure}[ht]
\centering
\begin{overpic}
[scale=0.67]{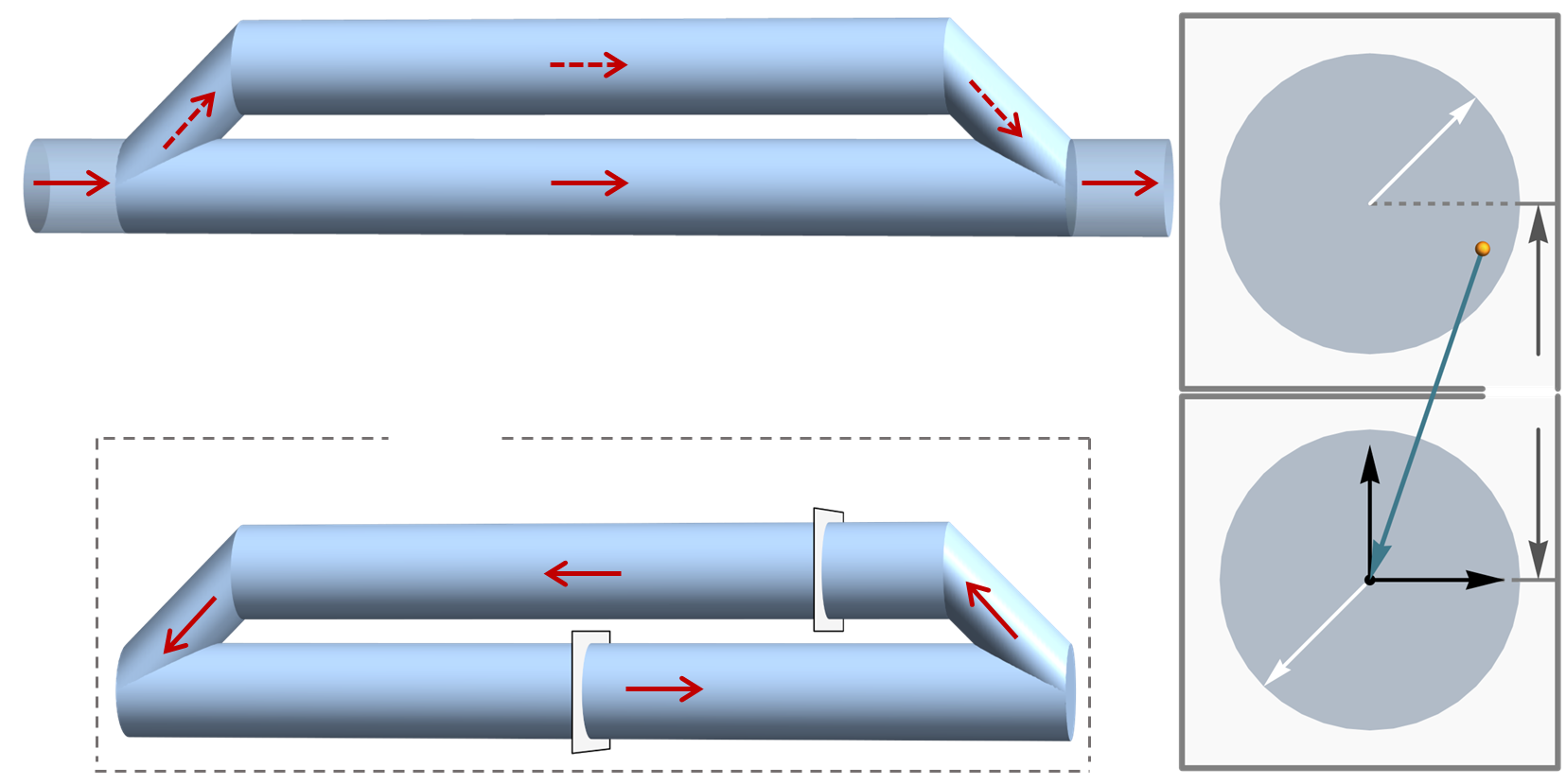}
\put(0,47){(a)}
\put(0,19){(b)}
\put(76,45){(c)}
\put(64,28){\color{black}\vector(1,0){10}}
\put(64,28){\color{black}\line(-1,-1){10.5}}
\put(92,21){$\mathbf{r}$}
\put(86.5,10){$\mathbf{x}$}
\put(91,33.5){$\mathbf{x^{\prime}}$}
\put(83.1,19.3){$x^{\prime}_2$}
\put(92.5,14.5){$x^{\prime}_1$}
\put(86.5,40.5){$r_{\mathrm{d}}$}
\put(81,11){$r_{\mathrm{d}}$}
\put(96.4,24.2){$w$}
\put(20,25.5){Equivalent current loop}
\put(26,21){$L^{\prime}_C$}
\put(29,37){$C_0$}
\put(4,42){$C_1$}
\put(70,42){$C_1$}
\put(8,46){$C_2$}
\put(29,45){$C_3$}
\put(65,45){$C_4$}
\put(7.5,36){\color{gray}\line(0,-1){29}}
\put(35,31){\color{gray}\vector(-1,0){27.5}}
\put(39,31){\color{gray}\vector(1,0){29}}
\put(68,36){\color{gray}\line(0,-1){29}}
\put(36,30){$2l$}
\put(15.2,21){\color{gray}\line(0,-1){4}}
\put(35.5,19){\color{gray}\vector(-1,0){20.3}}
\put(39.5,19){\color{gray}\line(1,0){14.8}}
\put(55.5,19){\color{gray}\vector(1,0){5}}
\put(60.5,21){\color{gray}\line(0,-1){4}}
\put(36,18){$2l^{\prime}$}
\put(29,5.2){$C_0$}
\put(8.5,14){$\bar{C}_2$}
\put(29,12.5){$\bar{C}_3$}
\put(63.5,14){$\bar{C}_4$}
\put(38,2){\color{black}\vector(1,0){36}}
\put(76,27.5){$S_{\bar{C}_3}$}
\put(76,3){$S_{C_0}$}
\end{overpic}\caption{\label{fig-6} (a) Alternative current loop with segments $C_2$, $C_3$, $C_4$, and fiber $C_1$; (b) an equivalent current loop $L^{\prime}_C$ constructed from segments $C_0$, $\bar{C}_4$, $\bar{C}_3$, and $\bar{C}_2$; and (c) two cross-sections $S_{\bar{C}_3}$ and $S_{C_0}$ in $C_0$, as shown on the right side.}
\end{figure}

Next, it will be demonstrated that only $C_0$ needs to be considered. The strategy is to construct an alternative loop to bypass $C_0$, which can be realized by a detour composed of segments $C_2$, $C_3$, and $C_4$, as depicted in Fig.~\ref{fig-6}(a). $C_3$ is designed as a shifted segment with the same radius $r_{\mathrm{d}}$ as $C_0$, a length $2l^{\prime}$ and an offset distance $w$$>$2$r_{\mathrm{d}}$. $C_2$ and $C_4$ are two oblique cylindrical segments designed to symmetrically connect $C_1$ and $C_3$ at matching angles. According to Eq.~(\ref{Appd-Zero-Contribution}), the connection of $C_1$, $C_2$, $C_3$, and $C_4$ forms a loop similar to $X$. Hence, the contribution from $C_1$ is the same as that from a wire segment formed by $C_2$, $C_3$, and $C_4$, but with the current reversed. In this way, the external displacement current at $\mathbf{x}$ produced by $L_C$ is equivalent to an effective current loop $L^{\prime}_C$, as shown in Fig.~\ref{fig-6}(b), where the segment labels displaying top-barred letters indicate a reversed current inside. $L^{\prime}_C$ is merely a small current loop used as a conceptual tool, rather than a physical circuit.

Considering a thinning process of these current loops, the impact of $\bar{C}_2$ and $\bar{C}_4$ becomes insignificant as the radius $r_{\mathrm{d}}$ approaches zero, since their sizes, proportional to $r_{\mathrm{d}}$, become negligible compared to $C_0$. As for $\bar{C}_3$, the integral involves the displacement $\mathbf{r}$ from a source point $\mathbf{x^{\prime}}$, as illustrated in a cross-sectional area $S_{\bar{C}_3}$, to the field point $\mathbf{x}$ at the center of the given cross-sectional area $S_{C_3}$ in $C_0$, i.e.,
\begin{subequations}\label{Appd-T3-r}
\begin{align}
\mathbf{r}=&-x^{\prime}_1\mathbf{\hat{e}}^{\prime}_1-(w+x^{\prime}_2)\mathbf{\hat{e}}^{\prime}_2-x^{\prime}_3\mathbf{\hat{e}}^{\prime}_3\\
r^2=&x^{\prime 2}_3+r^2_{\mathrm{p}}, r^2_{\mathrm{p}}=x^{\prime 2}_1+(w+x^{\prime}_2)^2\,,
\end{align}
\end{subequations}
which expresses the integral in $\bar{C}_3$ as
\begin{align}
\mathbf{J}^{\mathrm{ext}}_{\mathrm{D},3}
=&\,\mathbf{J}\!\int_{\bar{C}_3}\!\!\frac{2z^2-r^2_{\mathrm{p}}}{4\pi r^5}\mathrm{d}^3x^{\prime}
=\,\frac{\mathbf{J}}{2\pi}\!\int_{S_{\bar{C}_3}}\!\!\frac{l^{\prime}\mathrm{d}s^{\prime}}{(l^{\prime 2}+r^2_{\mathrm{p}})^{3/2}}\,,
\label{Appd-Zero-T3}
\end{align}
where $z$=$x^{\prime}_3$, $\mathrm{d}^3x^{\prime}$=$\mathrm{d}z\mathrm{d}s^{\prime}$, $\mathrm{d}s^{\prime}$ is an area element in $S_{\bar{C}_3}$, and $\mathbf{J}$ is $-J\mathbf{\hat{e}}^{\prime}_3$. It should be noted that the transverse components of displacement current in $\bar{C}_3$ are zero due to an odd mirror symmetry with the inversion of the coordinate $z$$(x^{\prime}_3)$. With the relation: $r_{\mathrm{p}}$$\geqslant$$w$$-$$r_{\mathrm{d}}$=$\sigma r_{\mathrm{d}}$ ($\sigma$$>$1), the magnitude of Eq.~(\ref{Appd-Zero-T3}) can be estimated as
\begin{align}
\left|\mathbf{J}^{\mathrm{ext}}_{\mathrm{D},3}\right|
<&\int_{S_{\bar{C}_3}}\!\frac{Jl^{\prime}\mathrm{d}s^{\prime}}{2\pi(l^{\prime 2}+\sigma^2r^2_{\mathrm{d}})^{3/2}}\nonumber\\
=&\frac{Jl^{\prime}r^2_{\mathrm{d}}}{2(l^{\prime 2}+\sigma^2r^2_{\mathrm{d}})^{3/2}}\rightarrow 0\,\,\mathrm{with}\,\,r_{\mathrm{d}}\rightarrow 0\,,
\label{Appd-Estimation-T3}
\end{align}
where $r_{\mathrm{d}}$$\rightarrow$0 represents the thinning process of $L^{\prime}_C$. Therefore, only the contribution of $C_0$ to the displacement current is nontrivial.

According to Fig.~\ref{fig-5}(c), the integration should be performed over the region $C_0/V_{\mathrm{d}}$. Using Eqs.~(\ref{Appd-T3-r}) with $w$=0 and $\mathbf{J}$=$J\mathbf{\hat{e}}^{\prime}_3$, the final result is found to be
\begin{align}
\mathbf{J}^{\mathrm{ext}}_{\mathrm{D}}
=&\,\mathbf{J}\!\int_{C_0/V_{\mathrm{d}}}\!\!\frac{2z^2-r^2_{\mathrm{p}}}{4\pi r^5}\mathrm{d}^3x^{\prime}
=\mathbf{J}\!\int_{\lambda}^l\!\mathrm{d}z\int_{S_{C_0}}\!\!\mathrm{d}s^{\prime}\frac{2z^2-r^2_{\mathrm{p}}}{2\pi r^5}\nonumber\\
=&\,\mathbf{J}\int_\lambda^l\!\frac{r^2_{\mathrm{d}}\mathrm{d}z}{(z^2+r^2_{\mathrm{d}})^{3/2}}=\left.\frac{\mathbf{J}z}{\sqrt{z^2+r^2_{\mathrm{d}}}}\right|^l_{z=\lambda}\rightarrow\mathbf{J}\,.
\label{Appd-Final-T0}
\end{align}
In the last limitation in Eq.~(\ref{Appd-Final-T0}), the thinning processes are represented by $\lambda$ and $r_{\mathrm{d}}$ approaching zero, respectively. This outcome is consistent with the result from Eq.~(\ref{Steady-Parallel-D-Current}).

\section{Direct causal sources}
The traditional cause-and-effect interpretation of Maxwell's equations originates mainly from the field retardation, whereas that of Newtonian mechanics is thought to be defined by Newton's first and second laws based on the force cause. These two approaches exhibit remarkable differences but also share a notable commonality. Here, I focus on a typical physical causality paradigm: determinism based on physical processes is governed by dynamical equations that progress along a definite time direction, irrespective of chaos or quantum uncertainty. A cause-and-effect event must occur within a physical process that \cite{Huang3}, especially in this context, involves wave propagation, field induction, and motion of charged matter. A physical process then means a continuous causal chain belonging to classical physics, inherently entailing the concept of local causality, as pointed out by Kinsler \cite{Kinsler1} and Savage \cite{Savage}. Kinsler proposed a mathematical foundation for a physical process using causal differential equations, in which the effect quantity is related to the cause quantity through the time derivative \cite{Kinsler1}. The effect depends on the history of the cause because it implies a step-like temporal factor, which is consistent with the Kramers-Kronig relations. In such processes, a continuous limit within a sequence of events creates an illusion of simultaneity, which, in principle, does not break the temporal causal structure.

As a result, the two electromagnetic laws become the fundamental dynamical equations, which can also be non-uniquely transformed into wave equations to describe emissions. However, the above causal dynamical model conflicts with the traditional interpretation of Maxwell's equations. On one hand, as a continuous physical process, the generation of a magnetic field is commonly described by the Amp\`{e}re-Maxwell law, which, in this causal reasoning, actually governs the evolution of the electric field rather than the magnetic field. Therefore, the evolution pattern of an electric field is quite understandable, much like how a force induces a change in a particle's velocity or how heat causes a change in temperature. In the light of the idea and development by Kinsler and Savage (K-S) \cite{Kinsler1,Savage,Kinsler2}, the generation of a vortex electric field can be expressed using the differential form of the Amp\`{e}re-Maxwell law
\begin{align}
\dot{\mathbf{D}}=\nabla\times\mathbf{H}-\mathbf{J}\,.
\label{Ampere-Maxwell-Time-Differential}
\end{align}
Obviously, there is no need to introduce the displacement current. Similarly, Faraday's law
\begin{align}
\dot{\mathbf{B}}=-\nabla\times\mathbf{E}
\label{Faraday-Time-Differential}
\end{align}
intends for the conversion of a magnetic field from a vortex electric field, the time integral of which then reveals a clear temporal sequence. As such, Savage stated that ``Faraday's law then says that the local cause of change in the magnetic field is spatial variation of the electric field, specifically its curl." \cite{Savage} This differs significantly from the common understanding of magnetic field generation by displacement current. 

On the other hand, as an inverse transformation (from vorticity to field itself) of Helmholtz decomposition or a solution of the Amp\`{e}re-Maxwell law, the instantaneous Biot-Savart integral (\ref{New-H}) presents a nonlocal simultaneous dependence on the external displacement current. This brings an issue of simultaneity in a local event involving field conversion and generation, which is generally characterized by spatial integral transformations; nevertheless, it appears to be incompatible with the time-asymmetric causal principle \cite{Frisch}. In a continuous field induction situation, the causal structure indicates a time-ordered event chain. Disproving simultaneity requires logically verifying any putative event chain, which is unattainable without physical processes. In practice, verifying the causal role of a field source is challenging due to the spatial dispersion and overlap among fields. Consequently, most conservative viewpoints stem from an exclusive reliance on the verification standard of retarded effect or from the criticism of the simultaneity relationship. Jefimenko pointed out that there is no time delay, therefore, no phase difference between the electric and magnetic fields in an electromagnetic wave \cite{Jefimenko2}. So, ``the two fields cannot cause each other (by the principle of causality, the ﬁelds should be out of phase if they create each other)." Similarly, Arthur stated that \cite{Arthur} ``the motion of charge causes both magnetic field and displacement current, and hence the displacement current cannot be the cause of the magnetic field," the context of which is a direct cross-product relationship between the magnetic field and the displacement current for a moving point charge. 

Counterintuitively, in the new interpretation view, the simultaneity problem no longer arises, since it is not the simultaneous Biot-Savart integral but rather Faraday's law (\ref{Faraday-Time-Differential}) that conveys a causal relation. In an electromagnetic induction experiment, the induction can be conceptually divided into three processes:
\begin{itemize}
\item Vorticity excitation: the vorticity of the electric field emerges or changes due to the motion of charges or variations in electric current;
\item Magnetic induction: the vorticity of the electric field subsequently induces a magnetic field or alters it, as described by Faraday's law (\ref{Faraday-Time-Differential});
\item Electric induction: the induced magnetic field and current causes variations of the electric field governed by the Amp\`{e}re-Maxwell law (\ref{Ampere-Maxwell-Time-Differential}).
\end{itemize}
Consider a common phenomenon: a magnet with a static magnetic field $\mathbf{B}$ moving slowly at a constant velocity $\bm{\upsilon}$. The notion that the magnetic field can be treated as a moving field with the same velocity, expressed as $\mathbf{B}_{\mathrm{m}}(\mathbf{x},t)$$\equiv$$\mathbf{B}(\mathbf{x}$$-$$\bm{\upsilon}t)$ with the approximation at low speeds, requires reinterpretation (not proof). In the context of the new causality framework, the motion in the first process induces an electric field $\mathbf{E}$$=$$-$$\bm{\upsilon}$$\times$$\mathbf{B}_{\mathrm{m}}$ with a nonzero vorticity excitation, $\nabla$$\times$$\mathbf{E}$$=$$\bm{\upsilon}$$\cdot$$\nabla\mathbf{B}_{\mathrm{m}}$, through the translational movement of complex intramolecular charges within the magnet. This is essentially driven by a relativistic effect. In the second process, the excited electric vorticity around the magnet continuously generates new magnetic fields that superimpose onto the existing field, which causes the magnetic field to move along with the magnet by obeying a vector advection equation: $\dot{\mathbf{B}}_{\mathrm{m}}$=$-$$\bm{\upsilon}$$\cdot$$\nabla\mathbf{B}_{\mathrm{m}}$. Similarly, the vorticity of the magnetic field in the third process makes the electric field to follow the magnet by the same advection equation.

Further, think about Faraday's electromagnetic induction by moving the magnet toward a closed coil. In the first process, the electromotive force generated in the coil is a natural consequence of the vortex electric field, as given by its loop integral. Therefore, the coil current is resulted from the electric force as usual. In the second process, on one hand, the electric vorticity generated by the moving magnet sustains $\mathbf{B}_{\mathrm{m}}$; on the other hand, the coil current induces new electric vorticity in space, thereby generating new magnetic field $\delta \mathbf{B}$ that opposes $\mathbf{B}_{\mathrm{m}}$ (Lenz's law). Next, in the third process, the magnetic vorticity $\nabla$$\times$$\delta \mathbf{B}$ suppresses the electric field in the coil through Eq.~(\ref{Ampere-Maxwell-Time-Differential}), which is just self-induction. 

In fact, a moving particle carrying a charge $q$ encompasses all the processes described above in consideration of relativistic effects. The electric and magnetic fields will be mutually induced by each other, causing them move together. Therefore, the causal interpretation of a general electromagnetic process can, in principle, be decomposed and reduced into the superposition of single particle cases.

This field-centric causal interpretation is self-consistent and offers an alternative perspective for understanding. The pedagogy behind the K-S thought aims not only to uncover the diversity of interpretations but also to provide another way to comprehend the world, enabling us to develop personalized physical views.


\begin{thebibliography}{99}
\bibitem{Huang1}
The total current is commonly considered as a ``mixing" of nonlocal displacement current and a localized, real current. The concept of total current has been rarely addressed in recent textbooks, primarily due to this mixing and its indefinite role \cite{Jackson1,Griffiths}. In terms of the development in this paper, it actually is a nonlocal field (external displacement current) rather than a mixing.
\bibitem{Huang2}
The intrinsic magnetism from atoms and subatomic particles can be attributed to quantized circulating currents, including semi-classical spin currents. Hence, the electric current is the only matter source of magnetic fields.
\bibitem{Huang3}
It is difficult to find an exact and universally comprehensive definition of the concept of a physical process. However, one description provided by AI is acceptable to some extent, which states: a physical process is a naturally occurring or artificially induced sequence of changes in the state or properties of matter, energy, or physical systems that are governed by the laws of physics.
\end{thebibliography}
\end{document}